\documentclass[man, natbib, apacite, noextraspace, floatsintext]{apa6} %

\usepackage{graphicx, subfigure}
\graphicspath{{images/}}
\usepackage{array} 
\usepackage{amsmath, amssymb}
\usepackage[utf8]{inputenc}
\usepackage{mathtools}
\usepackage{booktabs, caption} %
\usepackage{threeparttable} 
\usepackage{xcolor, soul}
\usepackage{setspace}
\newcommand{\lp}{\left(}
\newcommand{\rp}{\right)}
\newcommand{\lb}{\left[}
\newcommand{\rb}{\right]}

\newcommand{\lds}{, \, \ldots \, , \,}
\newcommand{\bs}{ \, | \,} 
\newcommand{\ds}{\displaystyle}

\definecolor{gray}{rgb}{0.5, 0.5, 0.5}
\definecolor{grn}{rgb}{0.0, 0.5, 0.0}
\everymath{\displaystyle}

\usepackage[pagewise]{lineno}

\newcolumntype{L}[1]{>{\raggedright\let\newline\\\arraybackslash\hspace{0pt}}m{#1}}
\newcolumntype{C}[1]{>{\centering\let\newline\\\arraybackslash\hspace{0pt}}m{#1}}
\newcolumntype{R}[1]{>{\raggedleft\let\newline\\\arraybackslash\hspace{0pt}}m{#1}}

\title{Markov Network for Modeling Local Item Dependence in Cognitively Diagnostic Classification Models}
\date{}

\author{Hyeon-Ah Kang$^{1}$ \\ 
Jingchen Liu$^{2}$ \\ 
Zhiliang Ying$^{2}$ 
}

\affiliation{$^{1}$University of Texas at Austin \\
$^{2}$Columbia University}

\shorttitle{Graphical Diagnostic Classification Model}

\abstract{The study presents an exploratory graphical modeling approach for evaluating local item dependency within cognitively diagnostic classification models (DCMs). Current approaches to modeling local dependence require known item structure and have limited utility when such information is not available. In this study, we propose an exploratory approach to modeling local dependence so that items’ own interactions can be revealed without dependency specification. The new framework is developed by integrating a Markov network into a generalized DCM. The framework unveils item interactions while performing regular cognitive diagnosis within a unified scheme. The inference on the model parameters is made on the regularized pseudo-likelihood and is implemented by an EM algorithm. Numerical experimentation from Monte Carlo simulation suggests that the proposed framework adequately recovers generating parameters and yields reliable standard error estimates. Compared with the regular DCM, the new model produced more accurate item parameter estimates as items exhibit local dependence. The study demonstrates application of the model using two real assessment data and discusses practical benefits of modeling local dependence.

\mbox{} 
}

\begin{document}
\maketitle

\setcounter{page}{1}

\setlength{\abovedisplayskip}{1pt}
\setlength{\belowdisplayskip}{1pt}

\section{1. Introduction}

A cognitively diagnostic classification model (DCM; \citeNP<e.g.,>{delaTorre_2011, Haertel_1989, Henson_Templin_Willse_2009, vonDavier_2005}) is a family of a latent class model that describes examinee’s latent status on a set of discrete cognitive attributes. The model provides detailed information about individual’s latent profile and has been used to assess students’ skill mastery and diagnose patients’ pathological symptoms. One of the assumptions of diagnostic modeling is local item independence---examinee’s response variables on the items must be conditionally independent given the examinee’s cognitive profile. Once the examinee’s latent status is taken into account, the response variables are no longer expected to correlate with each other.

While the local independence assumption provides an important foundation for cognitive diagnosis and measurement validity, it appears a strong assumption and is not always warranted in real settings. In real assessments, items often exhibit excessive correlation when they are presented in similar forms or on the same stimuli (e.g., reading passage, laboratory scenario). Items measuring similar contents or sharing the same words can also exhibit redundant correlation. The excess of inter-item dependency above and beyond the relation implied by the latent status can adversely affect measurement outcomes and must be identified for proper remediation.

While a number of approaches have been presented for addressing local item dependence~\cite<e.g.,>{Asparouhov_Muthen_2011, Hansen_2013, Lim_2022, Sha_2016, Zhan_etal_2018}, the existing approaches are mostly designed for confirmatory analysis and require pre-specification of dependent items. For example, a classical approach is to identify items that are suspected of local dependency and model redundant covariance through a random-effect term~\cite<e.g.,>{Hansen_2013, Zhan_etal_2018}. This approach however lends little utility when the dependency structure is not known and can lead to overfitting when not all the specified items exhibit local dependency. A less explicit way of modeling local dependence is to use Bayes prior. \citeA{Asparouhov_Muthen_2011} applied an informative prior to estimate inter-item covariance matrices when fitting a latent class model. This approach, although it does not require known item relationships, requires informative priors for estimating the covariance matrix, which can consequently affect the dependency evaluation and the performance of a model~\cite{Lee_etal_2020}. Additionally, the approach estimates covariance matrix for each latent class separately (i.e., conditional inter-item covariance matrix given each latent class) and can lead to conflicting conclusions when the estimated covariance matrices differ across the latent classes.

The purpose of this study is to explore an alternative modeling approach to evaluating local item dependence. We in particular seek for exploratory modeling such that items’ local interactions can be unveiled without prior specification. Specifically, our strategy is to integrate a Markov network into a DCM and explicitly model redundant correlation between the items during diagnostic modeling. The Markov network provides a graphical tool for delineating conditional dependence of random variables and can point to hidden features that underlie the dependent variables. When merged into a DCM, it can model local interplay between the items and can explain away residual covariance. In this study, we demonstrate formulation of the new model and present an inferential scheme that simultaneously optimizes for DCM and Markov item network.


Following sections provide details of the proposed framework. Section 2 introduces a generalized DCM that serves as a baseline of the new framework and presents an Ising model that provides Markov network for binary response variables. The baseline model is formulated in a generalized form so that it can accommodate various forms of item-attribute interactions. The section continues with the modeling framework to integrate the constituting models and formulate a network-integrated DCM. The new model is called a graphical diagnostic classification model (GDCM) with the graph intended to model items’ local interplay. The ensuing section presents an inferential scheme for GDCM, including the estimation of model parameters, standard errors, and examinees’ latent attribute profiles. The subsequent Sections 4 and 5 each verify performance of the estimation method and of GDCM through Monte Carlo simulation. Each simulation study examines the accuracy and precision of the estimation and the relative performance of GDCM to a regular DCM. Section 6 provides example applications of the model to real assessment data, one from an international educational assessment and the other from a personality inventory. Section 7 concludes the article with a summary of findings, practical implications, and directions for future research.

\section{2. Modeling Framework}

\subsection{2.1. Diagnostic Classification Model}

The study applies a generalized DCM to accommodate various cognitive operation rules (e.g., conjunctive, disjunctive, additive). Let $\boldsymbol{\alpha}_i = (1, \, \alpha_{i1}, \, \alpha_{i2} \lds \alpha_{iK}, \, \alpha_{i1} \alpha_{i2}, \cdots, $ $ \alpha_{i1} \alpha_{i2} \cdots \alpha_{iK})^\top$ denote examinee $i$’s latent cognitive status on a set of $K$ attributes. Mastery (or possession) of the individual attributes is coded by binary values---$\alpha_{ik} = 1$ if mastered; 0 otherwise ($k = 1 \lds K$). DCM assumes that the examinee’s latent attribute status is manifested by responses on the measurement items. Let $\boldsymbol{X}_i = (X_{i1} \lds X_{iJ})^\top$ denote the examinee’s response variables on $J$ test items. The probability of an item response is modeled by logistic regression:
\begin{linenomath*}
\begin{equation}
    p(X_{ij} = x_{ij} \bs \boldsymbol{\alpha}_i) = \frac{ \exp \lp x_{ij} \boldsymbol{\beta}_j^\top \boldsymbol{\alpha}_i \rp }{ 1 + \exp \lp \boldsymbol{\beta}_j^\top \boldsymbol{\alpha}_i \rp } \label{eq:irf}     
\end{equation}
\end{linenomath*}
with $x_{ij}$ ($\in \{0, \, 1\}$) denoting a realization of $X_{ij}$, and $\boldsymbol{\beta}_j = (\beta_{j0}, \, \beta_{j1} \lds \beta_{jK}, \, \beta_{j12} \lds $ $\beta_{j12\cdots K})^\top$ parameterizing the effects of $\boldsymbol{\alpha}$ on the logit. We assume that $\boldsymbol{\beta}_j$ is properly formulated to embody item’s loading on the attributes. For example, for an item that measures the first attribute only, $\boldsymbol{\beta}_j$ takes the form of $(\beta_{j0}, \, \beta_{j1}, \, \mathbf{0})^\top$ with $\beta_{j0} \in \mathbb{R}$ and $\beta_{j1} \in \mathbb{R}^+$. A matrix that compiles this item-attribute incidence is known as a $\boldsymbol{Q}$-matrix~\cite{Tatsuoka_1985}, $\boldsymbol{Q} = (\boldsymbol{q}_j^\top: \, j = 1 \lds J)$, where an entry vector $\boldsymbol{q}_j$ indicates presence and absence of item-attribute interaction---$q_{jk} = 0$ if present; 0 otherwise.

In settings where items are locally independent, examinee’s test performance can be summarized as 
\begin{linenomath*}
\begin{equation}
    p (\boldsymbol{X}_i = \boldsymbol{x}_i \bs \boldsymbol{\alpha}_i) = \prod_{j=1}^J \frac{ \exp \lp x_{ij} \boldsymbol{\beta}_j^\top \boldsymbol{\alpha}_i \rp }{ 1 + \exp \lp \boldsymbol{\beta}_j^\top \boldsymbol{\alpha}_i \rp } \label{eq:dcm}
\end{equation}
\end{linenomath*}
with $\boldsymbol{x}_i = (x_{i1} \lds x_{iJ})^\top$. This simplification provides a foundation for estimating the item and person parameters. However, as alluded to above, the local independence assumption is not always warranted and may be violated in real assessments. In this study, we present a flexible modeling framework that enables cognitive diagnosis in the presence of latent local item dependence.

\subsection{2.2. Markov Item Network}

Our strategy for modeling local dependence is to integrate a Markov network into a DCM so that items’ interrelationships can be modeled by a pairwise Markov network.\footnote{The study aims to model the weak form of local item dependence, i.e., pairwise dependence~\cite{McDonald_1979}. The pairwise independence implies higher-order independence (i.e., the strong form of local independence)~\cite{McDonald_Mok_1995, Stout_2002}.} The Markov network provides a succinct graphical summary of conditional dependence of random variables and can be easily integrated into a latent variable model to describe redundant covariance.

In the context of modeling item response covariance, items represent vertices of the network, $\mathbb{V} = \{1 \lds J\}$, and an edge between the vertices, $(j, \, j^\prime)$, models conditional dependence between the items.\footnote{The conditional (in)dependence is evaluated given other items on the test.} The strength of the dependence is characterized by a $J$-by-$J$ symmetric network matrix, $\boldsymbol{S} = (s_{jj^\prime}: j, \, j^\prime \in \mathbb{V})$, with entries taking zeros when items are conditionally independent and nonzero values when items are conditionally dependent. A set of item pairs with nonzero $s_{jj^\prime}$ entries then defines an edge set, $\mathbb{E} = \{(j, \, j^\prime): \, s_{jj^\prime} \ne 0 \} \subset \mathbb{V} \times \mathbb{V}$, containing locally dependent item pairs.

Given the network matrix, a joint probability distribution of the response variables is modeled as
\begin{linenomath*}
\begin{equation}
    p(\boldsymbol{X}_i = \boldsymbol{x}_i) = \frac{1}{z(\boldsymbol{S})} \exp \lp \frac{1}{2} \boldsymbol{x}_i^\top \boldsymbol{S} \boldsymbol{x}_i \rp \label{eq:im}
\end{equation}
\end{linenomath*}
with a normalizing constant, $z(\boldsymbol{S}) = \sum_{\boldsymbol{x} \in \{0, \, 1\}^J} \exp \lp \frac{1}{2} \boldsymbol{x}^\top \boldsymbol{S} \boldsymbol{x} \rp$. The formulation in \eqref{eq:im} is known as the Ising model~\cite{Ising_1925}. The model is a type of Markov random field that describes pairwise interactions between binary variables. A nonzero pattern in the network matrix shows local interactions between the items and the size of $s_{jj^\prime}$ models the strength of the interaction.

\subsection{2.3. Graphical Diagnostic Classification Model}

To combine the two models, we draw on a proportionality-invariant relationship. Let $\propto$ represent a proportional relationship that differs in a proportionality constant. The DCM \eqref{eq:dcm} can be then re-expressed as
\begin{linenomath*}
\begin{equation}
    p(\boldsymbol{X}_i = \boldsymbol{x}_i \bs \boldsymbol{\alpha}_i) \propto \exp \lp \boldsymbol{x}_i^\top \boldsymbol{B} \boldsymbol{\alpha}_i \rp \label{eq:dcm_kernel}
\end{equation}
\end{linenomath*}
with $\boldsymbol{B} = (\boldsymbol{\beta}_j^\top: \, j = 1 \lds J)$ containing all item parameter vectors. Similarly rewriting \eqref{eq:im}, we obtain
\begin{linenomath*}
\begin{equation}
    p(\boldsymbol{X}_i = \boldsymbol{x}_i) \propto \exp \lp \frac{1}{2} \boldsymbol{x}_i^\top \boldsymbol{S} \boldsymbol{x}_i \rp. \label{eq:im_kernel} 
\end{equation}
\end{linenomath*}
Both the equations differ from their original forms by the normalizing constants only.

Integrating the models, we obtain a model that describes the joint distribution of $\boldsymbol{X}_i$. Let $p(\boldsymbol{x})$ denote $p(\boldsymbol{X} = \boldsymbol{x})$. The same abbreviation is used to denote the probability measures hereafter. The kernel function of the conditional joint probability distribution of $\boldsymbol{x}_i$ given $\boldsymbol{\alpha}_i$ and item interplay is obtained as
\begin{linenomath*}
\begin{equation}
    p(\boldsymbol{x}_i \bs \boldsymbol{\alpha}_i, \, \boldsymbol{B}, \boldsymbol{S}) \propto \exp \lp \boldsymbol{x}_i^\top \boldsymbol{B} \boldsymbol{\alpha}_i + \frac{1}{2} \boldsymbol{x}_i^\top \boldsymbol{S} \boldsymbol{x}_i \rp. \label{eq:gdcm_kernel}
\end{equation}
\end{linenomath*}
Let $\pi_{\boldsymbol{\alpha}_i}$ denote the prior class probability of the attribute profile, $\boldsymbol{\alpha}_i$, and $\boldsymbol{\pi}$ contain all class probabilities, $\boldsymbol{\pi} = (\pi_{\boldsymbol{\alpha}_c}: \, c = 1 \lds 2^K)$. GDCM then defines the joint probability distribution of $(\boldsymbol{x}_i, \, \boldsymbol{\alpha}_i)$ as
\begin{linenomath*}
\begin{equation}
    p(\boldsymbol{x}_i, \, \boldsymbol{\alpha}_i \bs \boldsymbol{B}, \, \boldsymbol{S}, \, \boldsymbol{\pi}) = \frac{ \pi_{\boldsymbol{\alpha}_i} }{z (\boldsymbol{B}, \, \boldsymbol{S}, \, \boldsymbol{\pi})} \exp \lp \boldsymbol{x}_i^\top \boldsymbol{B} \boldsymbol{\alpha}_i + \frac{1}{2} \boldsymbol{x}_i^\top \boldsymbol{S} \boldsymbol{x}_i \rp \label{eq:gdcm}
\end{equation}
\end{linenomath*}
with a normalizing constant, $z (\boldsymbol{B}, \, \boldsymbol{S}, \, \boldsymbol{\pi}) = \sum_{\boldsymbol{\alpha} \in \{0, \, 1\}^K} \sum_{\boldsymbol{x} \in \{0, \, 1\}^J} \pi_{\boldsymbol{\alpha}} \exp \lp \boldsymbol{x}^\top \boldsymbol{B} \boldsymbol{\alpha} + \frac{1}{2} \boldsymbol{x}^\top \boldsymbol{S} \boldsymbol{x} \rp $. Again, the model is called a graphical DCM with the graph modeling items’ local interplay.

In modeling \eqref{eq:gdcm}, we assume that examinee’s attribute profile accounts for most response variance and only a small portion of variance results from item interactions. This assumption encourages sparsity of the network matrix such that most entries of $\boldsymbol{S}$ take zero values (i.e., local independence) and only a small set takes nonzero values. This assumption should be tenable in most assessment settings as it implies lack of test validity otherwise. Also relating to the parameterization of $\boldsymbol{S}$, we constrain diagonal entries at zero and assume that any main effects of the items can be traced to the interaction between the item and examinee’s attribute mastery status. This setup aligns with the convention in the Markov random fields that does not allow self-interaction. 

It follows from \eqref{eq:gdcm} that the conditional joint probability distribution of $\boldsymbol{x}_i$ given $\boldsymbol{\alpha}_i$ takes the form:
\begin{linenomath*}
\begin{equation}
    p(\boldsymbol{x}_i \bs \boldsymbol{\alpha}_i, \, \boldsymbol{B}, \, \tilde{\boldsymbol{S}}, \, \boldsymbol{\pi}) = \frac{ \exp \lp \boldsymbol{x}_i^\top \tilde{\boldsymbol{S}} \boldsymbol{x}_i \rp }{\ds \sum_{\boldsymbol{\alpha} \in \{0, \, 1\}^K} \sum_{\boldsymbol{x} \in \{0, \, 1\}^J} \pi_{\boldsymbol{\alpha}} \exp \lp \boldsymbol{x}^\top \tilde{\boldsymbol{S}} \boldsymbol{x} \rp } \label{eq:gdcm_joint}
\end{equation}
\end{linenomath*}
with $\tilde{\boldsymbol{S}} = \tilde{\boldsymbol{S}}(\boldsymbol{\alpha}, \, \boldsymbol{B}, \, \boldsymbol{S}) = (\tilde{s}_{jj^\prime}: \, j, \, j^\prime = 1 \lds J) = \begin{cases} \boldsymbol{\beta}_j^\top \boldsymbol{\alpha} & \; \, \text{if} \; j = j^\prime \\ s_{jj^\prime} & \; \, \text{if} \; j \ne j^\prime \end{cases}$. It can be further shown that the conditional probability of an item response variable follows logistic regression:
\begin{linenomath*}
\begin{equation}
    p(x_{ij} \bs \boldsymbol{\alpha}_i, \, \boldsymbol{x}_{i, \, -j}, \boldsymbol{B}, \, \boldsymbol{S}) = \frac{ \exp \lp x_{ij} \boldsymbol{\beta}_j^\top \boldsymbol{\alpha}_i + x_{ij} \boldsymbol{s}_{-j}^\top \boldsymbol{x}_{i, \, -j} \rp }{ 1 + \exp \lp \boldsymbol{\beta}_j^\top \boldsymbol{\alpha}_i + \boldsymbol{s}_{-j}^\top \boldsymbol{x}_{i, \, -j} \rp }, \label{eq:gdcm_cond}
\end{equation}
\end{linenomath*}
where $\boldsymbol{x}_{i, \, -j} = (x_{i1} \lds x_{i, \, j-1}, \, x_{i, \, j+1} \lds x_{iJ})$ is a vector of item responses that excludes the response on item, $j$.\footnote{An item subscript with a minus sign indicates an item set that excludes the corresponding item(s), e.g., $\boldsymbol{x}_{-j} = (x_j: \, j \in \mathbb{V} \setminus \{j\} \}$.} This notion helps develop an estimation scheme followed below.


\section{3. Pseudo-Likelihood Inference}

\subsection{3.1. Parameter Estimation}

\newpage

The GDCM contains three sets of model parameters, $\boldsymbol{\theta} = (\boldsymbol{B}, \, \boldsymbol{S}, \, \boldsymbol{\pi})$, and one incidental variable, $\boldsymbol{\alpha}$. For estimating the model parameters, an arrangement needs to be made to deal with the latent variable, $\boldsymbol{\alpha}$. In this study, we apply an expectation-maximization (EM) algorithm and obtain model parameter estimates from the mode of the marginalized log-likelihood. Let $\boldsymbol{\theta}^{(t)} = (\boldsymbol{B}^{(t)}, \, \boldsymbol{S}^{(t)}, \, \boldsymbol{\pi}^{(t)})$ denote a set of provisional estimates of the free parameters obtained at the $t$th iteration of EM. The algorithm then iteratively maximizes the expected log-likelihood of the complete-data given $\boldsymbol{\theta}^{(t)}$:
\begin{linenomath*}
\begin{equation}
    E \lb \log L (\boldsymbol{\theta}; \, \boldsymbol{x}_i, \, \boldsymbol{\alpha}) \, \Big| \, \boldsymbol{x}_i, \, \boldsymbol{\theta}^{(t)} \rb = \sum_{\boldsymbol{\alpha} \in \{0, \, 1\}^K} p(\boldsymbol{\alpha} \bs \boldsymbol{x}_i, \, \boldsymbol{\theta}^{(t)}) \, \log p(\boldsymbol{x}_i, \, \boldsymbol{\alpha} \bs \boldsymbol{\theta}), \label{eq:em}
\end{equation}
\end{linenomath*}
where $L (\boldsymbol{\theta}; \, \boldsymbol{x}_i, \, \boldsymbol{\alpha})$ gives complete-data likelihood of an examinee’s response data. The expectation is taken with respect to the posterior probability of $\boldsymbol{\alpha}$ given $(\boldsymbol{x}_i, \, \boldsymbol{\theta}^{(t)})$. 

A direct evaluation of $p(\boldsymbol{x}_i, \, \boldsymbol{\alpha} \bs \boldsymbol{\theta})$ requires computation of a normalizing constant over $2^J \times 2^K$ terms and incurs high computational cost. A viable solution is to use surrogate likelihood that gives a computationally affordable approximate. In this study, we apply pseudo-likelihood~\cite{Besag_1975} that simplifies the conjoint problem into a local neighbor problem. The pseudo-likelihood is computationally more workable and yields estimates that closely approximate maximum likelihood estimates for dyad problems~\cite{vanDuijn_etal_2009}. In the present setting, the pseudo-likelihood of an examinee’s complete data is obtained as
\begin{linenomath*}
\begin{equation}
    L (\boldsymbol{\theta}; \, \boldsymbol{x}_i, \, \boldsymbol{\alpha}) = p(\boldsymbol{x}_i, \, \boldsymbol{\alpha} \bs \boldsymbol{\theta}) \overset{\Delta}{=} \prod_{j=1}^J p(x_{ij}, \, \boldsymbol{\alpha} \bs \boldsymbol{x}_{i, \, -j}, \, \boldsymbol{\theta}) = \prod_{j=1}^J p(x_{ij} \bs \boldsymbol{\alpha}, \, \boldsymbol{x}_{i, \, -j}, \, \boldsymbol{\theta}) \, p(\boldsymbol{\alpha} \bs \boldsymbol{x}_{i, \, -j}, \, \boldsymbol{\theta}), \label{eq:pl}
\end{equation}
\end{linenomath*}
where $\overset{\Delta}{=}$ redefines the likelihood, and $\boldsymbol{x}_{i, \, -j} = (x_{i1} \lds x_{i, \, j-1}, \, x_{i, \, j+1} \lds x_{iJ})$ denotes the examinee $i$’s response vector that excludes the $j$th item. The two probability functions in the right most hand of \eqref{eq:pl} are obtained as
\begin{linenomath*}
\begin{equation*}
    p(x_{ij} \bs \boldsymbol{\alpha}, \, \boldsymbol{x}_{i, \, -j}, \, \boldsymbol{\theta}) = \phi_{ij}^{x_{ij}} \lp 1 - \phi_{ij} \rp^{1 - x_{ij}}
\end{equation*}
\end{linenomath*}
with $\phi_{ij} = \phi_{ij} (\boldsymbol{\alpha}) = p(X_{ij} = 1 \bs \boldsymbol{\alpha}, \, \boldsymbol{x}_{i, \, -j}, \, \boldsymbol{B}, \, \boldsymbol{S}) = \mathrm{logit}^{-1} \lp \boldsymbol{\beta}_j^\top \boldsymbol{\alpha}_i + \boldsymbol{s}_{-j}^\top \boldsymbol{x}_{i, \, -j} \rp$   (see \eqref{eq:gdcm_cond}), and
\begin{linenomath*}
\begin{equation*}
    p(\boldsymbol{\alpha} \bs \boldsymbol{x}_{i, \, -j}, \, \boldsymbol{B}, \, \boldsymbol{S}, \, \pi_{\boldsymbol{\alpha}}) = \frac{ \pi_{\boldsymbol{\alpha}} p( \boldsymbol{x}_{i, \, -j} \bs \boldsymbol{\alpha}, \, \boldsymbol{B}, \, \boldsymbol{S} )}{\ds \sum_{\boldsymbol{\alpha}_c \in \{0, \, 1\}^K} \pi_{\boldsymbol{\alpha}_c} p( \boldsymbol{x}_{i, \, -j} \bs \boldsymbol{\alpha}_c, \, \boldsymbol{B}, \, \boldsymbol{S} ) },
\end{equation*}
\end{linenomath*}
where $\boldsymbol{s}_{-j} = (s_{1j}, \, s_{2j} \lds s_{j-1, \, j}, \, s_{j+1, \, j} \lds s_{Jj})^\top$ denotes the $j$th column of $\boldsymbol{S}$ that excludes the $j$th entry, and $p( \boldsymbol{x}_{i, \, -j} \bs \boldsymbol{\alpha}, \, \boldsymbol{B}, \, \boldsymbol{S} )$ is evaluated by the pseudo-likelihood.

For a sample of $N$ examinees, the logarithm of the pseudo-likelihood is obtained as the sum of $N$ log-pseudo-likelihoods. Let $\boldsymbol{X} = (\boldsymbol{x}_i: \, i = 1 \lds N)$ and $\boldsymbol{A} = (\boldsymbol{\alpha}_i: \, i = 1, $ $\cdots, \, N)$ each denote the response data and attribute matrix of the calibration sample. The conditional expectation of $\log L(\boldsymbol{\theta}; \, \boldsymbol{X}, \, \boldsymbol{A})$ is then maximized at the M-step of the EM algorithm to attain an updated iterate. Specifically, the objective function that is maximized at the M-step is obtained as
\begin{linenomath*}
\begin{align}
    \mathcal{Q} (\boldsymbol{\theta} \bs \boldsymbol{\theta}^{(t)} ) &= E \lb \, \log L (\boldsymbol{\theta}: \, \boldsymbol{X}, \, \boldsymbol{A}) \, \Big| \, \boldsymbol{X}, \, \boldsymbol{\theta}^{(t)} \, \rb \notag \\ &= \sum_{i=1}^N \sum_{\boldsymbol{\alpha} \in \{0, \, 1\}^K} \lb \, p(\boldsymbol{\alpha} \bs \boldsymbol{x}_i, \, \boldsymbol{\theta}^{(t)}) \sum_{j=1}^J \big[ \log p(x_{ij} \bs \boldsymbol{\alpha}, \, \boldsymbol{x}_{i, \, -j}, \, \boldsymbol{\theta}) + \log p(\boldsymbol{\alpha} \bs \boldsymbol{x}_{i, \, -j}, \, \boldsymbol{\theta}) \big] \, \rb \label{eq:qfn}
\end{align}
\end{linenomath*}
given the current provisional estimate, $\boldsymbol{\theta}^{(t)}$. The E- and M-steps iterate alternatively until convergence. The final estimates of the free parameters are then obtained as
\begin{linenomath*}
\begin{equation}
    (\hat{\boldsymbol{B}}, \, \hat{\boldsymbol{S}}, \, \hat{\boldsymbol{\pi}}) = \mathrm{arg \; max} \; \, \log L (\boldsymbol{B}, \, \boldsymbol{S}, \, \boldsymbol{\pi}; \, \boldsymbol{X}, \, \boldsymbol{A}). \label{eq:est}
\end{equation}
\end{linenomath*}

\textbf{\textit{Standard error estimation.}} The pseudo-likelihood simplifies the $J$ conjoint problem into a dyad problem (i.e., an item vs. its neighbor) and may underestimate the uncertainty of parameter estimates. In this study, we investigate three standard error estimators and evaluate empirical fidelity of the error estimates. The estimators of choice are: (i) the Hessian, (ii) outer-product of gradient (OPG), and (iii) sandwich estimators. Let $l (\hat{\boldsymbol{\theta}})$ denote the logarithm of the pseudo-likelihood evaluated at the final estimates. Each estimator estimates covariance of the final estimates as
\begin{linenomath*}
\begin{flalign*}
    \hat{V}_H &= \lp - \nabla^2 l (\hat{\boldsymbol{\theta}}) \rp^{-1} \quad \text{(Hessian)} & \\
    \hat{V}_G &= \lp \nabla l (\hat{\boldsymbol{\theta}}) \cdot \nabla l (\hat{\boldsymbol{\theta}})^\top \rp^{-1} \quad \text{(OPG), and} & 
\end{flalign*}
\end{linenomath*}

\begin{linenomath*}
\begin{flalign*}
    \hat{V}_S &= \hat{V}_H ( \hat{V}_G^{-1} )^{-1} \hat{V}_H, \quad \text{(sandwich)} &
\end{flalign*}
\end{linenomath*}
where $\nabla$ denotes a graident. The standard error estimate is then obtained as a square root of the diagonal entries of the covariance matrix.

\textbf{\textit{Attribute classification.}} Solving the $\mathcal{Q}$-function \eqref{eq:qfn} for the model parameters readily produces posterior probabilities of $\boldsymbol{\alpha}$ as a by-product. One can make capital of this outcome to estimate examinee’s attribute mastery profile:
\begin{linenomath*}
\begin{equation}
    \hat{\boldsymbol{\alpha}}_i = \underset{ \boldsymbol{\alpha} \in \{0, \, 1\}^K }{\mathrm{arg \; max}} \; \, p(\boldsymbol{\alpha} \bs \boldsymbol{x}_i, \, \hat{\boldsymbol{B}}, \, \hat{\boldsymbol{S}}, \, \hat{\boldsymbol{\pi}}), \label{eq:map}
\end{equation}
\end{linenomath*}
where $\hat{\boldsymbol{\alpha}}_i$ gives a maximum a posteriori estimate.

\subsection{3.2. Computation}

The estimation problem \eqref{eq:est} involves simultaneous optimization of a multitude of parameters and is susceptible to nonconvergence. In this study, we apply a practicable algorithm, coordinate descent~\cite{Friedman_etal_2007, Fu_1998}, to solve the optimization.\footnote{In the present setting, the algorithm is more precisely referred to as coordinate ascent as the optimization is aimed to maximize likelihood. We retain the nomenclature as acknowledged to be in line with the literature.} The algorithm provides computational efficiency by solving for each coordinate separately and is known to achieve global optima after sufficient cyclic iterations~\cite{Friedman_etal_2010, Tseng_2001}. In the present setting, the algorithm can be implemented as follows.

\textbf{\textit{Network matrix}} ($\boldsymbol{S})$. From \eqref{eq:qfn}, the log-likelihood for estimating the network matrix can be simplified to
\begin{linenomath*}
\begin{equation}
    l_s = \sum_{i=1}^N \sum_{\boldsymbol{\alpha} \in \{0, \, 1\}^K} p(\boldsymbol{\alpha} \bs \boldsymbol{x}_i, \, \boldsymbol{\theta}^{(t)}) \sum_{j=1}^J \log p(x_{ij} \bs \boldsymbol{\alpha}, \, \boldsymbol{x}_{i, \, -j}, \, \boldsymbol{B}^{(t)}, \, \boldsymbol{S} ). \label{eq:lls}
\end{equation}
\end{linenomath*}
Since the conditional probability of $x_{ij}$, $p(x_{ij} \bs \boldsymbol{\alpha}, \, \boldsymbol{x}_{i, \, -j}, \, \boldsymbol{\theta})$, follows logistic regression (see \eqref{eq:gdcm_cond}), the maximization problem can be solved by iteratively reweighted least squares estimation. Applying the strategy of \citeA{Friedman_etal_2010}, we approximate the log-likelihood, $l_s$, by Taylor series expansion and search for a mode of $l_s$ at the local approximation. The quadratic approximation of the logistic log-likelihood is obtained as
\begin{linenomath*}
\begin{equation*}
    \log p(x_{ij} \bs \boldsymbol{\alpha}, \, \boldsymbol{x}_{i, \, -j}, \, \boldsymbol{B}^{(t)}, \, \boldsymbol{S}) \approx - \frac{1}{2} \omega_{ij}^{(t)} \lp \boldsymbol{\beta}_j^{(t)\top} \boldsymbol{\alpha} + \boldsymbol{s}_{-j}^\top \boldsymbol{x}_{i, \, -j} - y_{ij}^{(t)} \rp^2,
\end{equation*}
\end{linenomath*}
where
\begin{linenomath*}
\begin{flalign*}
    \omega_{ij}^{(t)} &= \phi_{ij}^{(t)} (1 - \phi_{ij}^{(t)}) & \\
    \phi_{ij}^{(t)} &= p(x_{ij} = 1 \bs \boldsymbol{\alpha}, \, \boldsymbol{x}_{i, \, -j}, \, \boldsymbol{\theta}^{(t)}) = \mathrm{logit}^{-1} ( \boldsymbol{\beta}_j^{(t)\top} \boldsymbol{\alpha} + \boldsymbol{s}_{-j}^{(t)\top} \boldsymbol{x}_{i, \, -j} ), \; \mathrm{and} & \\
    y_{ij}^{(t)} &= \boldsymbol{\beta}_j^{(t)\top} \boldsymbol{\alpha} + \boldsymbol{s}_{-j}^\top \boldsymbol{x}_{i, \, -j} + (\omega_{ij}^{(t)})^{-1} (x_{ij} - \phi_{ij}^{(t)}). &
\end{flalign*}
\end{linenomath*}

Recall that the network matrix is assumed to take sparse entries with most entries taking zeros and only a few entries taking nonzero values. To enforce sparsity, we penalize $L_1$ norm of the network matrix. Let $\lambda$ denote a regularization parameter that controls the sparsity (i.e., degree of freedom). The objective function to be maximized for $\boldsymbol{S}$ is then obtained as
\begin{linenomath*}
\begin{equation}
    \mathcal{Q}_s = \sum_{i=1}^N \sum_{\boldsymbol{\alpha} \in \{0, \, 1\}^K} p(\boldsymbol{\alpha} \bs \boldsymbol{x}_i, \, \boldsymbol{\theta}^{(t)}) \sum_{j=1}^J \lp - \frac{1}{2} \omega_{ij}^{(t)} \lp \boldsymbol{\beta}_j^{\top} \boldsymbol{\alpha} + \boldsymbol{s}_{-j}^\top \boldsymbol{x}_{i, \, -j} - y_{ij}^{(t)} \rp^2 \rp - N \lambda \sum_{ 1\le j < j^\prime \le J} | s_{jj^\prime} | \label{eq:Qs}
\end{equation}
\end{linenomath*}
with a symmetry constraint, $s_{jj^\prime} = s_{j^\prime j}$ $(1 \le j < j^\prime \le J)$. In \eqref{eq:Qs}, the $L_1$ penalty performs both the variable selection and regularization by concurrently determining nonzero entries and curbing extreme values. The estimation based on such algorithm is known to yield consistent estimates~\cite{Zhao_Yu_2007}.

Solving \eqref{eq:Qs} for $s_{jj^\prime}$, a least-squares estimate is obtained in the form of
\begin{linenomath*}
\begin{equation*}
    \hat{s}_{jj^\prime}^{(t+1)} = 
	\frac{ \ds		
		\frac{1}{N}
		\sum_{i=1}^N \sum_{\boldsymbol{\alpha} \in \{0, \, 1\}^K}
		p (\boldsymbol{\alpha} \bs \boldsymbol{x}_i, \, \boldsymbol{\theta}^{(t)} ) \,
		\omega_{ij}^{(t)}
		x_{ij^\prime} 
		r_{ijj^\prime}
            +
            (-1)^{\mathrm{sign}(r_{ijj^\prime})} \lambda
	}{ \ds
		\frac{1}{N} \sum_{i=1}^N \sum_{\boldsymbol{\alpha} \in \{0, \, 1\}^K}
		p (\boldsymbol{\alpha} \bs \boldsymbol{x}_i, \, \boldsymbol{\theta}^{(t)} ) \,
		\omega_{ij}^{(t)}
		x_{ij^\prime}^2
	}
\end{equation*}
\end{linenomath*}
with $r_{ijj^\prime} = y_{ij}^{(t)} - \boldsymbol{\beta}_j^{\top} \boldsymbol{\alpha} - \boldsymbol{s}_{-(j, \, j^\prime)}^\top \boldsymbol{x}_{i, \, - (j, \, j^\prime)}$ modeling the partial residual remained after regressing $y_{ij}^{(t)}$ on $(\boldsymbol{\alpha}, \, \boldsymbol{x}_{-(j, \, j^\prime)})$. Applying the soft-thresholding operator~\cite{Donoho_Johnstone_1994}, a new iterate for $\hat{s}_{jj^\prime}$ is obtained as
\begin{linenomath*}
\begin{equation*}
    \hat{s}_{jj^\prime}^{(t+1)} \longleftarrow 
	\frac{ \ds
		\mathcal{S} \lp 
		\frac{1}{N}
		\sum_{i=1}^N \sum_{\boldsymbol{\alpha} \in \{0, \, 1\}^K}
		p (\boldsymbol{\alpha} \bs \boldsymbol{x}_i, \, \boldsymbol{\theta}^{(t)} ) \,
		\omega_{ij}^{(t)}
		x_{ij^\prime} 
		r_{ijj^\prime}, \;
		\lambda
		\rp
	}{ \ds
		\frac{1}{N} \sum_{i=1}^N \sum_{\boldsymbol{\alpha} \in \{0, \, 1\}^K}
		p (\boldsymbol{\alpha} \bs \boldsymbol{x}_i, \, \boldsymbol{\theta}^{(t)} ) \,
		\omega_{ij}^{(t)}
		x_{ij^\prime}^2
	},
\end{equation*}
\end{linenomath*}
where $\mathcal{S}(z, \, \lambda)$ is the soft-thresholding operator:
\begin{linenomath*}
\begin{equation*}
    \mathcal{S} (z, \, \lambda) = \mathrm{sign} (z) (|z| - \lambda)_{+}
    =
    \begin{cases}
        z  - \lambda & \text{if } z > 0 \text{ and }  \lambda < |z| \\
        z + \lambda  & \text{if } z < 0 \text{ and }  \lambda < |z| \\
        0            & \text{if } \lambda \ge |z|.
    \end{cases}
\end{equation*}
\end{linenomath*}
 
\textbf{\textit{Tuning parameter}} ($\lambda$). In \eqref{eq:Qs}, the optimization is performed under a specific tuning parameter, and the choice of $\lambda$ will determine the success of regularization. Current practice of choosing $\lambda$ commonly considers two criteria for decision: the Bayesian information criterion~\cite<BIC;>{Schwarz_1978} and the Akaike information criterion~\cite<AIC;>{Akaike_1974}. BIC has been the common choice when consistent variable selection is desired. AIC or cross-validation~\cite{Craven_Wahba_1979} has been used when the efficiency is of interest. In this study, we follow the current convention and apply BIC to obtain a consistent estimate of the network matrix. Within each iteration, a tuning parameter is obtained as
\begin{linenomath*}
\begin{equation*}
    \lambda = \mathrm{arg \; max} \; \, \mathrm{BIC} (\mathcal{M}_{\lambda}) = \mathrm{arg \; max} \; \, 2 \log L_{\lambda} - |\mathcal{M}_{\lambda}| \log N
\end{equation*} 
\end{linenomath*}
from a plausible range with the model fit under $\lambda$, $\mathcal{M}_{\lambda}$, and the number of free model parameters, $|\mathcal{M}_{\lambda}|$.

\textbf{\textit{DCM item parameters}} ($\boldsymbol{B}$). The optimization for $\boldsymbol{\beta}$ can be achieved by the Newton’s method. Maximizing \eqref{eq:qfn} with respect to $\boldsymbol{\beta}_j$ is equivalent to maximizing
\begin{linenomath*}
\begin{equation*}
    l_{\beta} = \sum_{i=1}^N \sum_{\boldsymbol{\alpha} \in \{0, \, 1\}^K} p(\boldsymbol{\alpha} \bs \boldsymbol{x}_i, \, \boldsymbol{\theta}^{(t)}) \sum_{j=1}^J \big[ x_{ij} \log \phi_{ij} (\boldsymbol{\alpha}) + (1 - x_{ij}) \log (1 - \phi_{ij} (\boldsymbol{\alpha})) \big].  
\end{equation*}
\end{linenomath*}
An updated estimate is then obtained as
\begin{linenomath*}
\begin{equation*}
    \boldsymbol{\beta}_j^{(t+1)} \longleftarrow \boldsymbol{\beta}_j^{(t)} -(\boldsymbol{H}_j^{(t)})^{-1} \cdot \nabla_{j}^{(t)},
\end{equation*} 
\end{linenomath*}
where $\nabla_{j}^{(t)}$ and $\boldsymbol{H}_j^{(t)}$ each denote the gradient and Hessian of the log-likelihood, $l_{\beta}$, evaluated at $(\boldsymbol{\beta}_{j}^{(t)}, \, \boldsymbol{s}_j^{(t+1)})$. The elements of $\nabla_{j}^{(t)}$ and $\boldsymbol{H}_j^{(t)}$ are calculated as 
\begin{linenomath*}
\begin{flalign*}
    &\nabla_{jl}^{(t)} = \frac{\partial l_{\beta}}{\partial \beta_{jl}} = \sum_{i=1}^N \sum_{\boldsymbol{\alpha} \in \{0, \, 1\}^K} \alpha_l (x_{ij} - \phi_{ij}) \, p(\boldsymbol{\alpha} \bs \boldsymbol{x}_i, \, \boldsymbol{\theta}^{(t)}) \; \, \text{for } l = 1 \lds 2^K, \quad \text{and} & \\
    &\boldsymbol{H}_{j(l, \, l^\prime)}^{(t)} = \frac{\partial^2 l}{\partial \beta_{jl} \partial \beta_{jl^\prime} } = - \sum_{i=1}^N \sum_{\boldsymbol{\alpha} \in \{0, \, 1\}^K} \alpha_l \alpha_{l^\prime} \phi_{ij} (1 - \phi_{ij}) \, p(\boldsymbol{\alpha} \bs \boldsymbol{x}_i, \, \boldsymbol{\theta}^{(t)}) \; \, \text{for } l, \, l^\prime = 1 \lds 2^K.
\end{flalign*}
\end{linenomath*}

\textbf{\textit{Class prior}} ($\boldsymbol{\pi}$). The optimization of $\boldsymbol{\pi}$ is similarly performed on the simplified log-pseudo-likelihood. Observe that in \eqref{eq:qfn} the logistic log-likelihood does not contribute to the estimation of $\boldsymbol{\pi}$. Maximizing \eqref{eq:qfn} with respect to $\boldsymbol{\pi}$ is therefore equivalent to maximizing
\begin{linenomath*}
\begin{equation*}
    l_{\pi} = \sum_{i=1}^N \sum_{\boldsymbol{\alpha} \in \{0, \, 1\}^K} p(\boldsymbol{\alpha} \bs \boldsymbol{x}_i, \, \boldsymbol{\theta}^{(t)}) \sum_{j=1}^J \lb \, \log \pi_{\boldsymbol{\alpha}} - \log \lp \sum_{\boldsymbol{\alpha}_c \in \{0, \, 1\}^K} \pi_{\boldsymbol{\alpha}_c} p(\boldsymbol{x}_{i, \, -j} \bs \boldsymbol{\alpha}_c, \, \boldsymbol{B}, \, \boldsymbol{S}) \rp \, \rb.
\end{equation*}
\end{linenomath*}
A new estimate, $\pi_l^{(t+1)}$ $(l = 1 \lds 2^K)$, is then obtained as
\begin{linenomath*}
\begin{equation*}
    \pi_l^{(t+1)} \longleftarrow \pi_l^{(t)} - (H_l^{(t)})^{-1} \cdot \nabla_l^{(t)},
\end{equation*}
\end{linenomath*}
where $\nabla_l^{(t)}$ and $H_l^{(t)}$ each denote the gradient and Hessian of $l_{\pi}$ evaluated at $(\pi_l^{(t)}, \, \boldsymbol{B}^{(t+1)}, \, \boldsymbol{S}^{(t+1)})$:
\begin{linenomath*}
\begin{flalign*}
    &\nabla_{l}^{(t)} = \pi_l^{-1} \sum_{i=1}^N \sum_{j=1}^J \lb \, p(\boldsymbol{\alpha}_l \bs \boldsymbol{x}_i, \, \boldsymbol{\theta}^{(t)}) - p(\boldsymbol{\alpha}_l \bs \boldsymbol{x}_{i, \, -j}, \, \boldsymbol{B}, \, \boldsymbol{S}) \, \rb, \quad \text{and} &\\
    &H_l^{(t)} = - \pi_l^2 \sum_{i=1}^N \sum_{j=1}^J p(\boldsymbol{\alpha}_l \bs \boldsymbol{x}_i, \, \boldsymbol{\theta}^{(t)}) - p(\boldsymbol{\alpha}_l \bs \boldsymbol{x}_{i, \, -j}, \, \boldsymbol{B}, \, \boldsymbol{S})^2. &
\end{flalign*}
\end{linenomath*}

\textbf{\textit{Termination.}} The iteration continues until no significant improvement is made in all coordinate directions. The current study applies tolerance criteria of .0001 for absolute mean change and log-likelihood change, and .001 for absolute maximum change.

\section{4. Simulation Study I: Parameter Recovery}

The performance of the proposed methods was evaluated in a series of Monte Carlo simulation. We first evaluated performance of the estimation algorithm in recovering the model parameters and estimating the standard errors. We then evaluated relative performance of GDCM to a regular DCM under different degrees of local item dependence. Sections 4 and 5 present each study.

\subsection{4.1. Design}

\textbf{\textit{Factors.}} The first study was performed under two-factorial design: (i) the frequency \\

\noindent
of local dependence (no, dyad, and triad dependence) and (ii) calibration sample size ($N =$ 500, 1000, and 3000). A precursory simulation study suggested that the number of attributes has marginal impact on the recovery of the model parameters and it was fixed at a constant value, $K = 4$. The test length similarly affected only the running time and was fixed at a moderate value of $J = 30$.

\begin{figure}[pt!]
    \begin{center}
    \subfigure[Null (Local Independence)]{
        \includegraphics[scale=0.7]{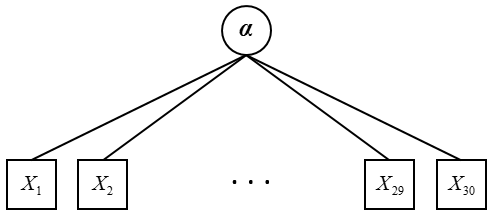} 
    } \\
    \vspace{20pt}
    \subfigure[Dyad Local Dependence]{
        \includegraphics[scale=0.7]{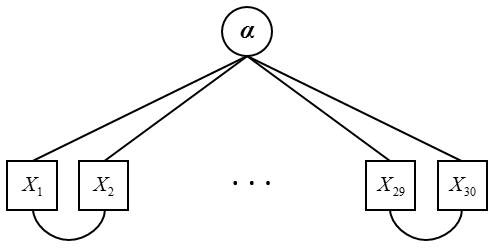} 
    } \\
    \vspace{20pt}
    \subfigure[Triad Local Dependence]{
        \includegraphics[scale=0.7]{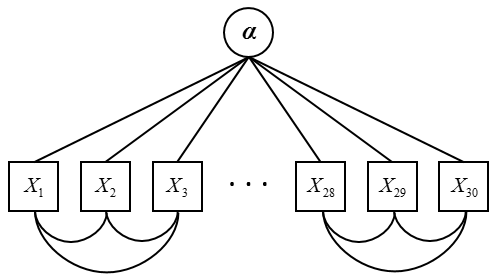}
    }
    \end{center} 
    \caption{Graphical Representation of Item Network}
    \label{fig:fig1_S}
\end{figure}

\textbf{\textit{Item network.}} The local item dependence was simulated for three scenarios: (i) the null case of local independence, (ii) dyad dependence where pairs of items exhibit extra  dependence, and (iii) triad dependence where triplets of items display redundant dependence (see Figure~\ref{fig:fig1_S} for graphical representation of each scenario). The null case assumes that all items are locally independent and have void network. The dyad condition assumes that adjacent item pairs (e.g., $\{1, \, 2\}, \, \{3, \, 4\} \lds \{29, \, 30\}$) are interrelated and have nonzero $s_{j,\, j+1}$ values at $j = 1, \, 3 \lds 29$. The triad dependence similarly assumes that adjacent item cliques (e.g., $\{1, \, 2, \, 3\}, \, \{4, \, 5, \, 6 \} \lds \{28, \, 29, \, 30\}$) have interaction with nonzero network entries (i.e., $s_{j,\,j+1}$, $s_{j,\,j+2}$, and $s_{j+1,\,j+2}$ nonzeros at $j = 1, \, 4 \lds 28$). For all nonzero entries, the $s$ values were randomly sampled from a standard normal distribution.

\textbf{\textit{DCM parameters.}} The measurement items were simulated assuming conjunctive item-attribute interaction. We first created an item-attribute incidence matrix, $\boldsymbol{Q} = (\boldsymbol{q}_j^\top: \, j = 1 \lds J)$, by randomly sampling $\boldsymbol{q}_j$ from $\{0, \, 1\}^K \setminus \mathbf{0}$. While sampling, a minimum constraint was imposed so that a test contains at least three unidimensional items measuring single attributes. This constraint ensures completeness of $\boldsymbol{Q}$ and identification of possible attribute profiles~\cite{Kohn_Chiu_2017}. The item-level parameters followed the formulation of the Deterministic Input Noisy Output ``AND'' Gate~\cite<DINA;>{Haertel_1989, Junker_Sijtsma_2001}. We randomly sampled guessing ($g_j$) and slipping ($s_j$) parameters from a uniform distribution, $U(.05, \, .2)$, emulating items of medium quality. Each parameter has a one-to-one correspondence with $\boldsymbol{\beta}$ on the generalized form, i.e., $g_j = \mathrm{logit}^{-1}(\beta_{j0})$ and $s_j = 1 - \mathrm{logit}^{-1} (\beta_{j0} + \beta_{j12 \cdots K_j})$, and this relationship was exploited when evaluating the parameter recovery on the generating metric. The other DCM structural parameter, $\boldsymbol{\pi}$, was generated from a uniform distribution with a constant probability, $\pi_l = 1/2^K$ $(l = 1 \lds 2^K)$, so that attributes have equal likelihood of mastery.

\textbf{\textit{Data generation.}} As the model parameters were obtained from the above setting, sample attribute profiles and response data were created as follows. The attribute mastery profiles were drawn from a multinomial distribution with .5 mastery probability (i.e., a sample with uniform class probabilities). The response data were generated from a Bernoulli distribution applying \eqref{eq:gdcm_cond}. In principle, response variables ought to be sampled from the joint probability distribution following \eqref{eq:gdcm_joint}. Doing so however requires calculation of normalizing constants and causes computational overload. In this study, we instead applied a Gibbs sampler~\cite{Lee_Hastie_2014} to simulate response data that approximate \eqref{eq:gdcm_joint}. Let $(x_1^{(t)} \lds x_{J}^{(t)})$ denote a set of draws at the $t$th sampling. A next response iterate is then sampled from a Bernoulli distribution with a success probability,
\begin{linenomath*}
\begin{equation*}
    \phi_{ij}^{(t+1)} = p(X_{ij}^{(t+1)} = 1 \bs \boldsymbol{\alpha}_i, \, \boldsymbol{x}_{i, \, -j}^{(t)}, \, \boldsymbol{B}, \, \boldsymbol{S}) = \mathrm{logit}^{-1} \lp \boldsymbol{\beta}_j^\top \boldsymbol{\alpha}_i + \boldsymbol{s}_{-j}^\top \boldsymbol{x}_{i, \, -j}^{(t)} \rp,
\end{equation*}
\end{linenomath*}
where $\boldsymbol{x}_{i, \, -j}^{(t)}$ contains the $t$th iterate response vector that excludes the $j$th response. To ensure that data follow an appropriate stationary distribution, we used draws after a burn-in period of 300 iterates as the final response data.

\textbf{\textit{Estimation.}} As response data were created, we fit GDCM following the procedures outlined in Section 3. The estimation was performed in nested loops with the outer loop implementing the EM algorithm and the inner loop performing the coordinate descent. The iteration continued until the iterates converge within the tolerance criteria (e.g., maximum change = .001, mean parameter change = .0001). 

\textbf{\textit{Replication.}} All simulation conditions were repeated 100 times with unique generating parameters and data sets, and results are summarized by taking the average and standard deviation (SD) of the evaluation statistics.

\subsection{4.2. Evaluation}

Simulation outcomes were evaluated in two aspects: the parameter recovery and precision of standard error estimates. The accuracy of the parameter estimates was evaluated by bias and root mean squared error (RMSE). Each statistic was calculated as
\begin{linenomath*}
\begin{equation*}
    \mathrm{Bias} = \hat{\theta} - \theta \quad \text{and} \quad \mathrm{RMSE} = \sqrt{\frac{1}{M} \sum_{m=1}^M (\hat{\theta}_m - \theta_m)^2},
\end{equation*}
\end{linenomath*}
where $\theta$ denotes a free parameter, $\hat{\theta}$ gives the corresponding estimate, and $m$ $(= 1 \lds M)$ denotes the number of congeneric parameters within each replication. The evaluation statistics were then averaged across the replications to obtain summary statistics.

The precision of standard error estimates was evaluated by comparing theoretic values to empirical estimates. Within each simulation condition, we randomly sampled one replication and repeated data generation and parameter estimation 100 times to obtain empirical standard error estimates. The empirical estimate was obtained as the standard deviation of the point estimates obtained across the replications. We then evaluated fidelity of the theoretical estimates (i.e., Hessian, OPG, sandwich) to the empirical estimates through average absolute distance.

\subsection{4.3. Results}

Table~\ref{tab:tab1_gdcm_pr} reports bias and RMSE of the GDCM parameter estimates. The diagnostic model item parameters $(\boldsymbol{\beta})$ were reported separately for the intercept and slope as they differ in the parameter domains and showed distinct bias patterns. The class prior estimates $(\hat{\boldsymbol{\pi}})$ showed almost zero bias and RMSEs (.000 and .012 each on average), and the results were omitted for brevity. The values reported in Table~\ref{tab:tab1_gdcm_pr} on the whole suggest that the estimation performed reasonably well. The network estimates $(\hat{s})$ showed minimal bias and RMSEs (.000 and .189 on average). The item parameter estimates $(\hat{\beta})$ entailed somewhat larger estimation error but the magnitude was kept within a tolerable range (average bias = -.019, RMSE = .197). 

\begin{table}[pt!]
\centering
\caption{Recovery of the GDCM Parameters}
\setlength\tabcolsep{5pt}
\label{tab:tab1_gdcm_pr}
\vspace{-0.5em}

\begin{threeparttable}
{\small
\renewcommand{\arraystretch}{0.8}
\begin{tabular}{L{1.5cm} L{1.0cm} R{1.2cm} R{1.2cm} R{1.2cm} R{1.2cm} R{1.2cm} R{1.2cm}}
\toprule
    &   &\multicolumn{3}{c}{Bias} &\multicolumn{3}{c}{RMSE} \\ \cmidrule(l){3-5} \cmidrule(l){6-8}
LD	&$N$	&$s$	&$\beta_0$	&$\beta_k$	&$s$	&$\beta_0$	&$\beta_k$ \\
\midrule
Null	&500	&-.001	&-.001	&.036	&.035	&.203	&.092 \\
	&1000	&.000	&.008	&.020	&.022	&.140	&.066 \\
	&3000	&-.001	&.008	&.006	&.011	&.081	&.037 \\
Dyad	&500	&.000	&.008	&-.023	&.227	&.356	&.118 \\
	&1000	&.000	&.005	&-.043	&.218	&.306	&.096 \\
	&3000	&.000	&.010	&-.055	&.212	&.267	&.081 \\
Triad	&500	&-.001	&.067	&-.163	&.332	&.485	&.141 \\
	&1000	&.000	&.070	&-.175	&.325	&.440	&.125 \\
	&3000	&.000	&.066	&-.189	&.316	&.403	&.112 \\
\bottomrule
\end{tabular}
}

\vspace{0.5em}
\begin{tablenotes}
    \linespread{1} \footnotesize
    \item \textit{Note}. LD: Local dependence structure. $N$: Calibration sample size. RMSE: Root mean squared error. $s$: Item network entry. $\beta_0$: Item intercept parameter. $\beta_k$: Item slope parameter. 
\end{tablenotes}
\end{threeparttable}
\vspace{-1.5em}
\end{table}

Table~\ref{tab:tab1_gdcm_pr} also shows that the estimation performed expectably across the design variables. As calibration samples enlarged, the estimation demonstrated more accurate recovery of the model parameters. The frequent occurrence of local dependence led to increasing RMSEs and bias but the estimation errors were kept reasonably small. One visible pattern in the bias results deserves a comment. As items display more frequent local dependence, the intercept and slope estimates $(\hat{\beta}_0, \, \hat{\beta}_k)$ showed seemingly visible directional bias. Close inspection of the estimates revealed that the direction was related to the sign of $s_{jj^\prime}$. The positive local item dependence tended to lead to overestimated intercepts and consequently underestimated slopes. The negative local dependence led to underestimated intercepts and overestimated slopes. The present simulation tended to have slightly more frequent positive local dependence, and it led to the systematic trend in the bias statistics as reported.

In Table~\ref{tab:tab2_gdcm_se}, we report standard error estimates of the item parameter estimates obtained from the asymptotic estimators. For comparison, empirical estimates attained from the 100 repetitions are presented (i.e., SD of point estimates across the replicates). The reported values suggest that the estimators overall well approximated the uncertainty. The average absolute difference between the theoretic and empirical estimates equaled .020, suggesting reasonable precision. Among the three estimators, the OPG estimator demonstrated the closest adherence to the empirical estimates (average distance = .005), followed by the Hessian and sandwich estimators (.008 and .013 each). It is however to be noted that the OPG estimator, though showed highest fidelity, occasionally yielded abnormally large standard errors in $\hat{\beta}_k$, with the abnormality consequently affecting the precision of the sandwich estimator. Taken all together, our evaluation across the simulations suggested that the Hessian estimator performs most stably and reliably, and it appears the most sensible choice for estimating the standard errors. Given this observation, we apply the Hessian estimator for reporting standard errors in the ensuing analysis.

\begin{table}[pt!]
\centering
\caption{Standard Error Estimation}
\setlength\tabcolsep{5pt}
\label{tab:tab2_gdcm_se}
\vspace{-0.5em}

\begin{threeparttable}
{\small
\renewcommand{\arraystretch}{0.8}
\begin{tabular}{L{1.5cm} L{1.0cm} R{0.8cm} R{0.8cm} R{0.8cm} R{0.8cm} R{0.8cm} R{0.8cm} R{0.8cm} R{0.8cm}}
\toprule
    &   &\multicolumn{2}{c}{Hessian} &\multicolumn{2}{c}{OPG} &\multicolumn{2}{c}{Sandwich} &\multicolumn{2}{c}{Empirical} \\ \cmidrule(l){3-4} \cmidrule(l){5-6} \cmidrule(l){7-8} \cmidrule(l){9-10}
LD	&$N$	&$\beta_0$	&$\beta_k$	&$\beta_0$	&$\beta_k$ &$\beta_0$	&$\beta_k$	&$\beta_0$	&$\beta_k$ \\
\midrule
Null	&500	&.186	&.368	&.203	&.395	&.171	&.344	&.201	&.380 \\
	&1000	&.126	&.255	&.136	&.268	&.117	&.243	&.140	&.264 \\
	&3000	&.074	&.144	&.081	&.153	&.068	&.136	&.079	&.152 \\
Dyad	&500	&.188	&.383	&.202	&.406	&.175	&.362	&.211	&.365 \\
	&1000	&.130	&.275	&.142	&.290	&.120	&.260	&.152	&.265 \\
	&3000	&.074	&.159	&.081	&.168	&.067	&.150	&.086	&.160 \\
Triad	&500	&.208	&.312	&.231	&.339	&.188	&.288	&.246	&.342 \\
	&1000	&.148	&.229	&.162	&.246	&.136	&.213	&.173	&.235 \\
	&3000	&.084	&.129	&.092	&.139	&.076	&.119	&.107	&.140 \\
\bottomrule
\end{tabular}
}

\vspace{0.5em}
\begin{tablenotes}
    \linespread{1} \footnotesize
    \item \textit{Note}. LD: Local dependence structure. $N$: Calibration sample size. $\beta_0$: Item intercept parameter. $\beta_k$: Item slope parameter. Hessian: Hessian estimator. OPG: Outer-product of the gradient. Sandwich: Sandwich estimator. Empirical: Empirical standard error obtained as SD of 100 replicate point estimates. The results for the theoretic estimates were obtained from one randomly-sampled replication in each condition. The empirical estimates were obtained from the 100 repetitions of the selected replication in each condition.
\end{tablenotes}
\end{threeparttable}
\end{table}

\section{5. Simulation Study II: Model Performance}

As we verify the accuracy of the estimation routine, we performed a second simulation study to evaluate performance of GDCM. We in particular examined relative performance of GDCM to a regular DCM to gauge advantage of modeling item network. 

\subsection{5.1. Design}

The design variables were set similarly to Study I except for the item network and the measurement model. The network matrices followed the same structure but were conditioned at constant values to examine the impact of different degrees of local dependence. Specifically, the edgewise parameters of the interacting items were set at $s_{jj^\prime} = .5$ and 1 to simulate moderate and large positive local dependence.\footnote{The values were chosen based on the setting of the existing literature~\cite<e.g.,>{Hofling_Tibshirani_2009, Guo_etal_2010}.}  The other entries of the network matrix (i.e., node-wise parameters, $s_{jj^\prime}$ of locally independent items), as well as the other parameters of DCM (i.e., $K$, $\boldsymbol{\beta}$, $\boldsymbol{\pi}$), were set similarly to Study I. As response data were obtained from the new set of parameters, we applied both the GDCM and DCM to examine relative performance. GDCM was fit applying the pseudo-likelihood estimator and DCM was fit constraining the item network at a null matrix within the pseudo-likelihood estimation. Although DCM can be fit by standard likelihood or Bayesian estimators, the study controlled the estimation method to the same scheme so that the models can be evaluated under the same settings.

\subsection{5.2. Evaluation}

The simulation outcomes were evaluated on similar criteria with Study I. We examined accuracy of the model parameter estimates based on bias, RMSE and standard error. Together with these criteria, we also examined sensitivity and false alarm rates of the Markov network in signaling the local dependence. Each criterion was calculated as
\begin{linenomath*}
\begin{align*}
    &\text{Sensitivity} = \frac{\ds \sum_{j < j^\prime} \mathcal{I} (\hat{s}_{jj^\prime} \ne 0, \, s_{jj^\prime} \ne 0)}{\ds \sum_{j < j^\prime} \mathcal{I} (s_{jj^\prime} \ne 0) }, \; \text{and} \; \, \\
    &\text{False positive rate} = \frac{\ds \sum_{j < j^\prime} \mathcal{I} (\hat{s}_{jj^\prime} \ne 0, \, s_{jj^\prime} = 0)}{\ds \sum_{j < j^\prime} \mathcal{I} (s_{jj^\prime} = 0) },
\end{align*}
\end{linenomath*}
where $\mathcal{I}(\cdot)$ denotes the indicator function. Each criterion can be seen as evaluating the proportion of nonzero edges that were correctly or falsely identified within the estimated network. 

In addition to the model parameter recovery, we also examined accuracy of the models in estimating the examinees’ attribute profiles. Estimating model parameters based on \eqref{eq:qfn} naturally produces posterior probability distributions of $\boldsymbol{\alpha}$ for each response pattern, and examinees’ attribute mastery profiles can be estimated as the mode of the posterior probability (see \eqref{eq:map}). In this study, the accuracy of the attribute estimate was evaluated by pattern- and attribute-wise agreement rate with the true cognitive profile:
\begin{linenomath*}
\begin{align*}
    &\text{Pattern-wise agreement rate (PAR)} = \sum_{i=1}^N \frac{\mathcal{I} (\hat{\boldsymbol{\alpha}}_i = \boldsymbol{\alpha}_i) }{N}, \text{and} \\
    &\text{Attribute-wise agreement rate (AAR)} = \sum_{i=1}^N \sum_{k=1}^K \frac{\mathcal{I}(\hat{\alpha}_{ik} = \alpha_{ik})}{NK},
\end{align*}
\end{linenomath*}
with $\hat{\boldsymbol{\alpha}}_i = (\hat{\alpha}_{i1} \lds \hat{\alpha}_{iK})$ and $\boldsymbol{\alpha}_i = (\alpha_{i1} \lds \alpha_{iK})$ each denoting the estimated and true attribute profiles of examinee $i$.

In evaluating the above criterion statistics, we performed analysis of variance to examine the proportion of variance associated with the choice of a model. The analysis results are presented with partial $\eta^2$ $(= \mathrm{SS}_{\mathrm{effect}} / (\mathrm{SS}_{\mathrm{effect}} + \mathrm{SS}_{\mathrm{error}})$) as appropriate. Below presents evaluation results that showed the most distinction between the two models. Other results that are not reported can be found from Appendix. 

\subsection{5.3. Results}

In Table~\ref{tab:tab3_gdcm_in} we report criterion statistics evaluated for $\hat{\boldsymbol{S}}$. All results were obtained from the GDCM that explicitly modeled the item network. The reported values suggest that the estimation adequately identified the underlying network. The interaction estimates ($\hat{s}_{jj^\prime}$) showed average bias of -.001 and RMSE of .023 when items were locally independent. As items begin to exhibit interdependency, the estimates entailed greater error, but the overall magnitude was kept reasonably small with the average bias of -.051 and RMSE of .171. The trends in the alarm rates were generally consistent with expectations. The stronger the interaction and the larger the sample, the higher the sensitivity. The false positive rates expectedly decreased as the local dependence occurred to a stronger degree and more frequently. We note that even when the items were locally independent, GDCM maintained false alarm rates at the nominal level (.051 on average), showing no significant cost of overfitting.

\begin{table}[pt!]
\centering
\caption{Recovery of Item Network}
\setlength\tabcolsep{5pt}
\label{tab:tab3_gdcm_in}
\vspace{-0.5em}

\begin{threeparttable}
{\small
\renewcommand{\arraystretch}{0.8}
\begin{tabular}{L{1.0cm} L{1.0cm} R{0.9cm} R{0.9cm} R{0.9cm} R{0.9cm} R{0.9cm} R{0.9cm} R{0.9cm} R{0.9cm} R{0.9cm} R{0.9cm} R{0.9cm}}
\toprule
$s_{jj^\prime}$ &$N$   &Bias	&{\scriptsize RMSE}	&FPR	&Bias	&{\scriptsize RMSE}	&Sen	&FPR	&Bias	&{\scriptsize RMSE}	&Sen	&FPR \\ \cline{1-13}
    &   &\multicolumn{3}{c}{LI $(s_{jj^\prime} =0)$} &\multicolumn{4}{c}{Dyad Dependence} &\multicolumn{4}{c}{Triad Dependence} \\ \cmidrule(l){3-5} \cmidrule(l){6-9} \cmidrule(l){10-13}
.5	&500	&-.001	&.035	&.051	&-.032	&.127	&.269	&.044	&-.061	&.176	&.282	&.040 \\
    &1000	&.000	&.022	&.052	&-.030	&.121	&.411	&.042	&-.059	&.169	&.428	&.032 \\
    &3000	&-.001	&.011	&.052	&-.027	&.109	&.717	&.035	&-.055	&.156	&.671	&.026 \\
1.0	&500	&	    &       &       &-.051	&.214	&.710	&.034	&-.093	&.276	&.781	&.016 \\
    &1000	&       &       &       &-.044	&.187	&.843	&.027	&-.075	&.227	&.918	&.017 \\
    &3000	&       &       &       &-.033	&.143	&.963	&.021	&-.048	&.148	&.991	&.018 \\
\bottomrule
\end{tabular}
}

\vspace{0.5em}
\begin{tablenotes}
    \linespread{1} \footnotesize
    \item \textit{Note}. $s_{jj^\prime}$: Item network entry for locally dependent items (.5: Moderate dependence, 1.0: Large dependence). $N$: Calibration sample size. RMSE: Root mean squared error. Sen: Sensitivity. FPR: False positive rate. LI: Local independence.
\end{tablenotes}
\end{threeparttable}
\end{table}

\begin{table}[pb!]
\centering
\caption{Recovery of Item Parameters: Comparison between GDCM and DCM}
\setlength\tabcolsep{5pt}
\label{tab:tab4_compare_ipar}
\vspace{-0.5em}

\begin{threeparttable}
{\small
\renewcommand{\arraystretch}{0.8}
\begin{tabular}{L{1.0cm} L{1.0cm} L{1.0cm} R{0.9cm} R{0.9cm} R{0.9cm} R{0.9cm} R{0.9cm} R{0.9cm} R{0.9cm} R{0.9cm}}
\toprule
    &   &   &\multicolumn{4}{c}{Bias}   &\multicolumn{4}{c}{RMSE} \\ \cmidrule(l){4-7} \cmidrule(l){8-11}
    &   &   &\multicolumn{2}{c}{GDCM}   &\multicolumn{2}{c}{DCM} &\multicolumn{2}{c}{GDCM}   &\multicolumn{2}{c}{DCM} \\ \cmidrule(l){4-5} \cmidrule(l){6-7} \cmidrule(l){8-9} \cmidrule(l){10-11}
LD	&$s_{jj^\prime}$	&$N$	&$g$	&$s$  	&$g$	&$s$	&$g$	&$s$	&$g$	&$s$ \\
\midrule
Null	&0	&500	&.001	&.000	&.001	&-.001	&.021	&.031	&.020	&.029 \\
	&	&1000	&.001	&-.001	&.000	&-.001	&.015	&.021	&.014	&.021 \\
	&	&3000	&.001	&-.001	&.001	&.000	&.008	&.012	&.008	&.012 \\
Dyad	&.5	&500	&.018	&-.029	&.020	&-.031	&.031	&.041	&.031	&.041 \\
	&	&1000	&.017	&-.029	&.021	&-.031	&.025	&.036	&.028	&.038 \\
	&	&3000	&.015	&-.026	&.021	&-.031	&.020	&.031	&.025	&.035 \\
	&1.0&500	&.039	&-.039	&.053	&-.051	&.049	&.052	&.064	&.059 \\
	&	&1000	&.033	&-.036	&.054	&-.052	&.041	&.045	&.063	&.058 \\
	&	&3000	&.024	&-.030	&.054	&-.053	&.029	&.036	&.062	&.057 \\
Triad	&.5	&500	&.045	&-.046	&.051	&-.052	&.056	&.056	&.062	&.059 \\
	&	&1000	&.043	&-.047	&.052	&-.053	&.051	&.053	&.061	&.059 \\
	& 	&3000	&.039	&-.045	&.052	&-.053	&.046	&.049	&.059	&.057 \\
	&1.0&500	&.100	&-.060	&.152	&-.080	&.115	&.072	&.166	&.087 \\
	&	&1000	&.080	&-.054	&.153	&-.082	&.092	&.064	&.167	&.088 \\
	& 	&3000	&.049	&-.042	&.153	&-.083	&.057	&.049	&.166	&.089 \\
\bottomrule
\end{tabular}
}

\vspace{0.5em}
\begin{tablenotes}
    \linespread{1} \footnotesize
    \item \textit{Note}. LD: Type of local dependence. $s_{jj^\prime}$: Item network entry for locally dependent items (.5: Moderate dependence, 1.0: Large dependence). $N$: Calibration sample size. RMSE: Root mean squared error. $g$: Guessing parameter. $s$: Slipping parameter.
\end{tablenotes}
\end{threeparttable}
\vspace{-1.5em}
\end{table}

Table~\ref{tab:tab4_compare_ipar} reports bias and RMSEs of the item parameter estimates obtained from the GDCM and DCM. The statistics are reported on the scale of the generating parameters (i.e., guessing, slipping) to gauge the impact on the attribute recovery. The reported values suggest that the two models showed essentially no difference when items were locally independent. As items begin to interact, GDCM showed clear advantage in improving the bias and RMSE (average $\eta^2 = .125$; $p < .001$). The improvement in the estimation was greater as the dependence occurs more frequently and more strongly and as the samples become larger.

The trends within each evaluated model were generally consistent with expectations. As the frequency and intensity of the dependence escalate, both models entailed increasing estimation error. The trends related to the sample size were somewhat different between the models. In GDCM, increasing the sample size constantly led to improved estimation while it made little difference in DCM. It seemed that when structural local dependence is present between the items, DCM hardly profits from the large samples and is not able to mitigate bias in the parameter estimation.

Again, we mention that the systematic bias pattern in the guessing and slipping estimates was chiefly due to the setting of the current simulation. Since items were simulated to display positive local dependence, the kernel of the item response function (i.e., \eqref{eq:gdcm_kernel}) became inflated, and it led to overestimated $\boldsymbol{\beta}$s and consequently a systematic bias in $g$ and $s$. When items were simulated to take interaction in any direction, no significant systematic bias was present.

In addition to the recovery of the model parameters, we also examined standard error of the item parameter estimates and accuracy of the attribute profile classification. The two models on the whole showed comparable performance in estimating the standard errors and began to show palpable differences as items exhibit strong or frequent interactions. The classification outcomes showed virtually no difference and remained steadily robust across the simulation conditions. See Appendix for detailed results.

\section{6. Real Data Application}

As we verify the performance of GDCM, we applied the model to real assessment data to examine practical relevance. Two secondary data were adopted for experimental application—the data from (i) the Trends in International Mathematics and Science Study (TIMSS) and (ii) the Eysenck Personality Questionnaire-Revised (EPQ-R). TIMSS gives an example of a scholastic cognitive assessment. EPQ-R gives an example of a personality inventory. Each data were chosen to investigate the functioning of GDCM under the partially-known and unknown item structure. Both the data are publically accessible via online repositories and can be analyzed retrospectively.

\subsection{6.1. TIMSS}

\textbf{\textit{Data.}} The TIMSS data were obtained from the 2015 Grade 4 Mathematics Assessment. Among the released data, the study examined the data from Booklet 6. The booklet contained the most testlet items and it was reckoned that it can help evaluate performance of GDCM under the known item structure. For calibration, we extracted data from the US sample. The extracted data contained $N = 684$ students’ responses to $J = 29$ items that measured three content domains (Number, Geometric Shapes \& Measures, Data Display). The $\boldsymbol{Q}$-matrix was constructed based on the released item specification.

\textbf{\textit{Analysis.}} As the calibration response data were obtained, we fit both the GDCM and DCM applying the pseudo-likelihood estimator similarly to Simulation Study II. The fitted outcomes were evaluated for relative model goodness-of-fit and similitude of parameter estimates. For evaluating model fit, we performed nonparametric bootstrapping with 500 resamples. We randomly sampled data from the calibration data and evaluated log-likelihood of the model parameter values estimated from the original data. Let $\hat{\boldsymbol{\theta}}$ denote the set of parameter values being evaluated. The log-likelihood of $\hat{\boldsymbol{\theta}}$ given the bootstrapped data, $\boldsymbol{X}^{(b)}$, is evaluated as 
\begin{linenomath*}
\begin{align*}
    l(\hat{\boldsymbol{\theta}}; \, \boldsymbol{X}^{(b)}) &= \log L(\hat{\boldsymbol{B}}, \, \hat{\boldsymbol{S}}, \, \hat{\boldsymbol{\pi}}; \, \boldsymbol{X}^{(b)}) 
\end{align*}
\end{linenomath*}
\begin{linenomath*}
\begin{align*}
    \qquad \qquad \qquad &= \sum_{i=1}^N \log \lp \sum_{\boldsymbol{\alpha} \in \{0, \, 1\}^K} p(\boldsymbol{x}_i^{(b)}, \, \boldsymbol{\alpha} \bs \hat{\boldsymbol{B}}, \, \hat{\boldsymbol{S}}, \, \hat{\boldsymbol{\pi}} ) \rp \\
    &= \sum_{i=1}^N \log \lp \sum_{\boldsymbol{\alpha} \in \{0, \, 1\}^K} \pi_{\boldsymbol{\alpha}} \exp \lp  \boldsymbol{x}_i^{(b)\top} \hat{\boldsymbol{B}} \boldsymbol{\alpha} + \frac{1}{2} \boldsymbol{x}_i^{(b)\top} \hat{\boldsymbol{S}} \boldsymbol{x}_i^{(b)} \rp \rp - \log z(\hat{\boldsymbol{B}}, \, \hat{\boldsymbol{S}}, \, \hat{\boldsymbol{\pi}}).
\end{align*}
\end{linenomath*}
Observe that the logarithm of the normalizing constant remains the same across the bootstrap replicates. The empirical sampling distribution of the log-likelihood can be thus constructed on the un-normalized metric discounting the last term. We apply this approach to evaluate relative likelihood of the observed estimates, $l^{\mathrm{obs}} = l(\hat{\boldsymbol{\theta}}; \, \boldsymbol{X})$, in the bootstrapped data. 

\begin{figure}[pb!]
    \vspace{-1em}
    \begin{center}
    \includegraphics{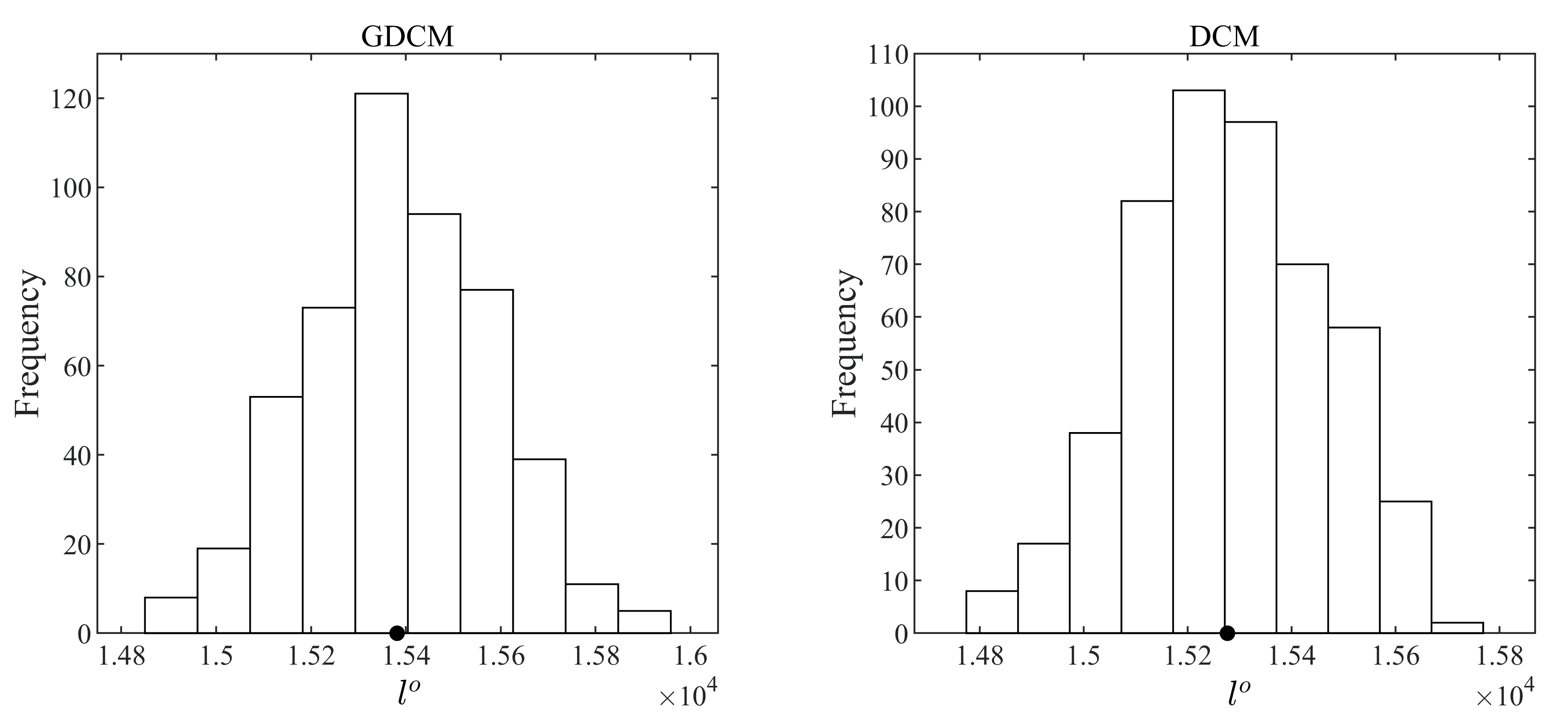}
    \end{center} 
    \vspace{-1em}
    \caption{TIMSS: Model Fit Evaluation}
    \label{fig:fig2_timss_fit}

    \vspace{0.5em}
    {\setstretch{0.5} \footnotesize \textit{Note}. $l^o$: Un-normalized log-likelihood of the model fit to the original calibration data. The empirical sampling distributions were obtained from 500 bootstrap resamples.}
\end{figure}

\textbf{\textit{Results.}} Figure~\ref{fig:fig2_timss_fit} presents empirical sampling distributions of the un-normalized log-likelihoods obtained from the bootstrap resamples. The log-likelihoods of the original calibration data were indicated to demonstrate the relative likeliness of the models. As can be noted, the two models showed acceptable fitness to the observed data. The cumulative probability of the un-normalized log-likelihood equaled .496 and .506 in each model, indicating no significant model misfit. We also mention that a similar pattern was observed when evaluated on the logarithm of the pseudo-likelihood (Equation \eqref{eq:pl}). Although the order of $p$-values differed (.224 (GDCM), .286 (DCM)), the two models uniformly showed no significant evidence of misfit.

\begin{figure}[pb!]
    \vspace{-1em}
    \begin{center}
    \includegraphics[scale=0.85]{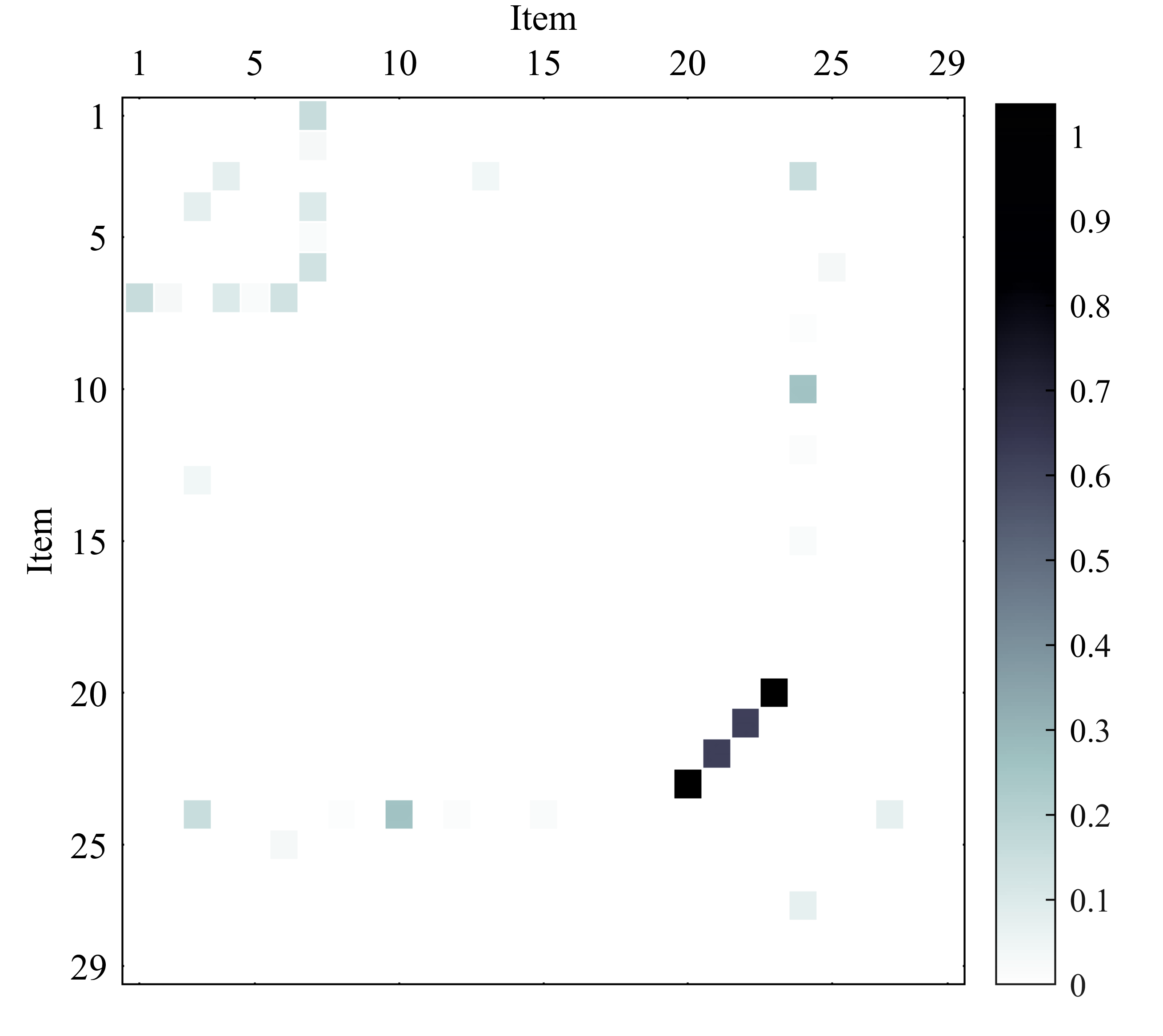}
    \end{center} 
    \vspace{-1em}
    \caption{TIMSS: Item Network}
    \label{fig:fig3_timss_heatmap}

    \vspace{0.5em}
    {\setstretch{0.5} \footnotesize \textit{Note}. The heat map is plotted for the absolute values of $\hat{s}_{jj^\prime}$. The darker the color, the stronger the interaction. Items (6, 25), (20, 21, 22, 23), (28, 29), respectively, were administered in testlets.}
\end{figure}

In Figure~\ref{fig:fig3_timss_heatmap}, we present a heat map of the item network estimated from the GDCM (see Table~\ref{tab:atab3_timss_ipair} for a complete list of locally dependent item pairs). The figure shows that items mostly remained locally independent and only a small set of items exhibited dependence (4.19\% of item pairs). The strongest interactions appeared between the items that were administered in testlets (e.g., $\hat{s}_{20, \, 23} = 1.037$, $\hat{s}_{21, \, 22} = .641$). Standalone items that were not tethered to testlets also showed local interactions with other items (e.g., Items 7, 24), suggesting probable existence of common stimuli that lead to extra correlation. One interesting pattern revealed by the heat map was that not all testlet items were disposed to local dependence. This finding runs contrary to the customary practice of modeling testlet items and illuminates potential functioning of GDCM. The model reveals surface local dependence of items and can circumvent overfitting of saturated testlet models.

\begin{figure}[pt!]
    \begin{center}
    \includegraphics[scale=0.9]{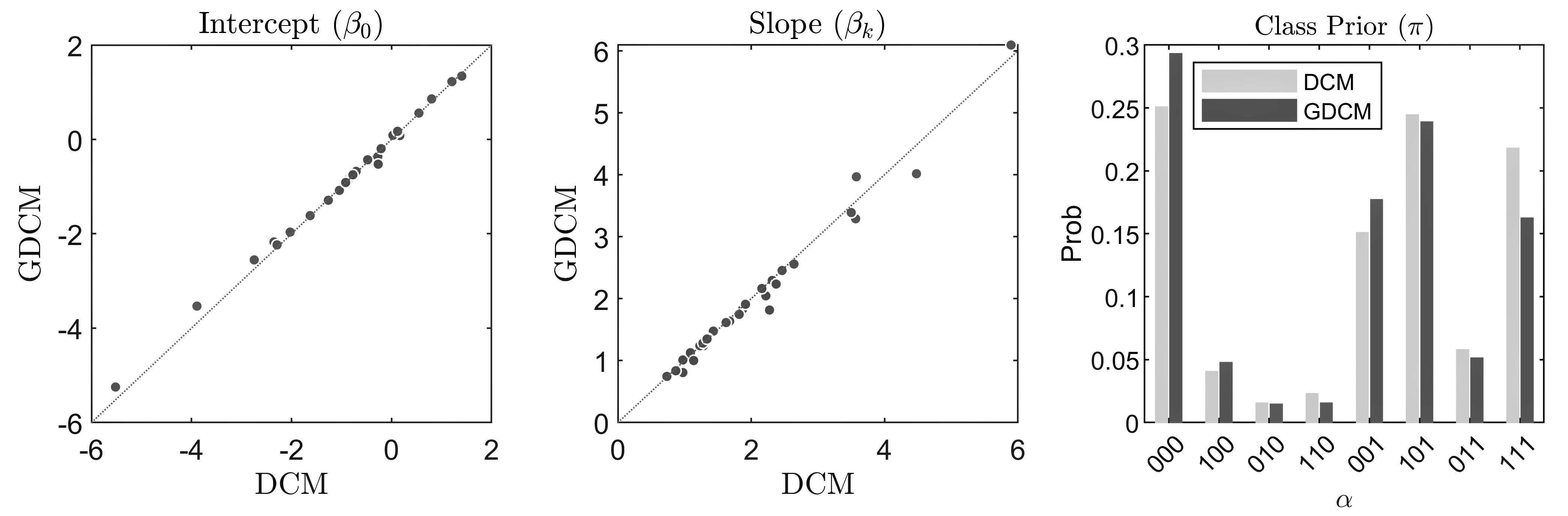}
    \end{center} 
    \vspace{-1em}
    \caption{TIMSS: Diagnostic Model Parameter Estimates}
    \label{fig:fig4_timss_ipar}

    \vspace{0.5em}
    {\setstretch{0.5} \footnotesize \textit{Note}. $\beta_0$: Item intercept parameter. $\beta_k$: Item slope parameter. $\boldsymbol{\pi}$: Latent class prior.}
\end{figure}

Presented in Figure~\ref{fig:fig4_timss_ipar} are diagnostic model parameter estimates obtained from the two models (see Table~\ref{tab:atab4_timss_ipar} for detailed results). The two models overall showed high consistency in the estimates. The item parameter values showed average correlation of .994, and the class prior estimates showed correlation of .964. While no distinct pattern could be identified in the item parameter values, we found that the differences in the item parameter values partially led to the discrete pattern in the class prior estimates. In Figure~\ref{fig:fig4_timss_ipar}, it is observed that DCM yielded a higher probability estimate for the full mastery group and a lower probability for the nonmastery group compared to GDCM. While the exact class distribution cannot be verified, we surmise that the unheeded local item dependence likely led to the overprediction of mastery and underprediction of nonmastery. Across the empirical analysis of real data, a similar pattern was observed with the DCM tending to overpredict mastery whenever the models disagree in the diagnostic model parameter values.

\subsection{6.2. EPQ-R}

\textbf{\textit{Data.}} The second example data, EPQ-R, was obtained from the cross-cultural validation study of \citeA{Eysenck_Barret_2013}. We obtained calibration data from the UK female sample that contained $N = 824$ responses. The questionnaire included 36 items measuring three dimensions of personal temperament---Psychoticism (P), Extraversion (E), and Neuroticism (N)---and 12 additional items on the Lie scale (L) measuring consistency in responses. For the analysis in this study, we examined the three scales that measured substantive constructs. We note that the instrument was originally intended to draw continuous scales. In this study, we aim to draw discrete attribute profiles indicating presence and absence of temperament. Similar practice has been applied in the DCM literature to diagnose pathological gamblers~\cite{Templin_Henson_2006} or to evaluate language proficiency~\cite{Jang_2009, Roussos_etal_2010}.

\textbf{\textit{Analysis.}} The analysis followed the same courses with the preceding analysis.

\textbf{\textit{Results.}} The two models showed approximately similar goodness-of-fit. GDCM showed a $p$-value of .504 and DCM of .464 when evaluated on the un-normalized log-likelihood; and .024 and .076 each when evaluated on the log-pseudo-likelihood. On the whole, we found that the models were not untenable and showed tolerable fitness to the observed data.

\begin{figure}[pt!]
    \vspace{-1em}
    \begin{center}
    \includegraphics[scale=0.85]{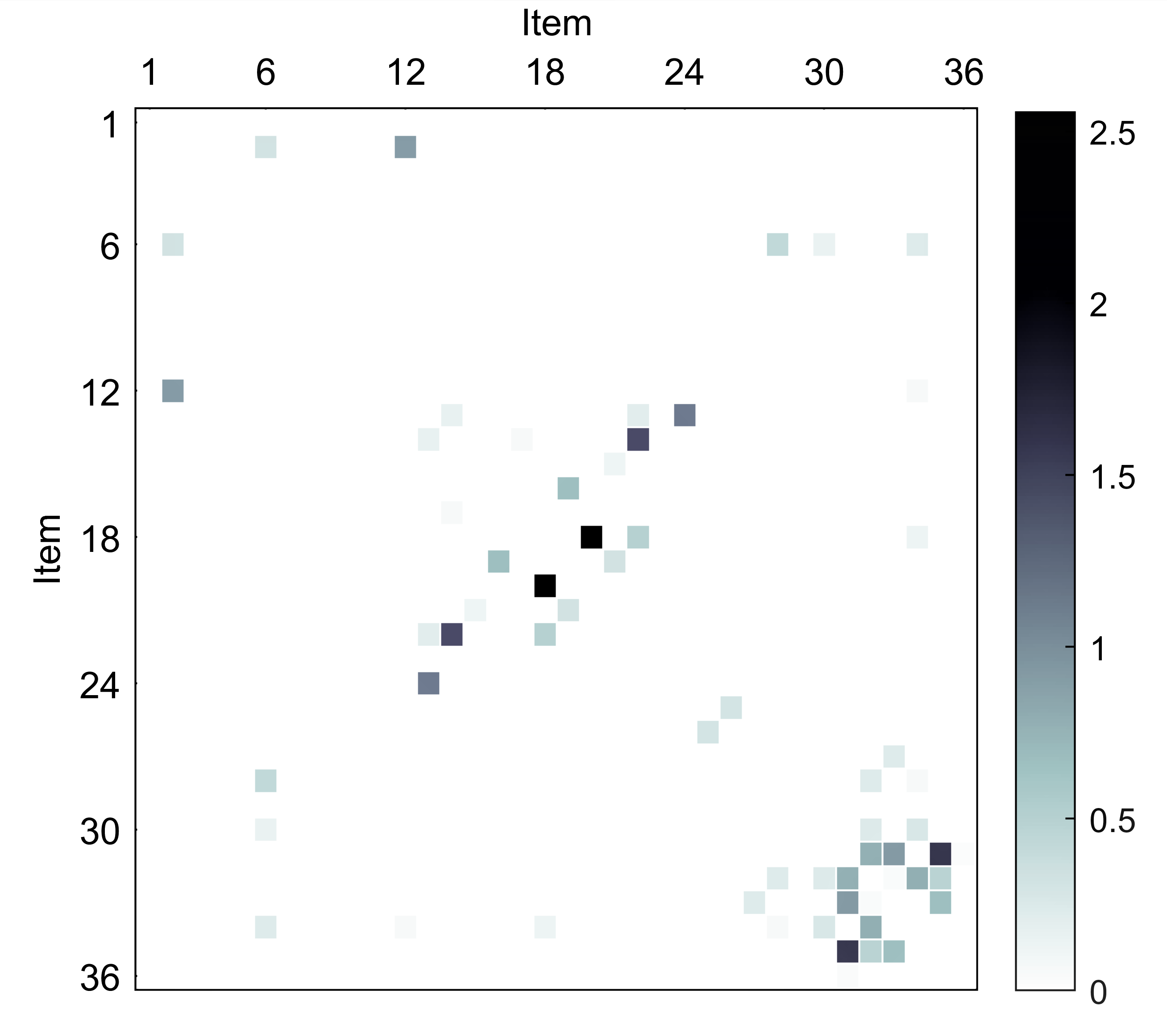}
    \end{center} 
    \vspace{-1em}
    \caption{EPQ-R: Item Network}
    \label{fig:fig5_epq_heatmap}

    \vspace{0.5em}
    {\setstretch{0.5} \footnotesize \textit{Note}. The heat map is plotted for the absolute values of $\hat{s}_{jj^\prime}$. The darker the color, the stronger the interaction. Items from 1 to 12 measure P (P); those from 13 to 24 measure E (E), and those from 25 to 36 measure N (N).}
    \vspace{-1em}    
\end{figure}

Figure~\ref{fig:fig5_epq_heatmap} presents a heat map of the item network estimated from GDCM (see Table~\ref{tab:atab5_epq_ipair} for a complete list of locally dependent item pairs). The network showed sparsity of 95.22\% with 31 nonzero edges (4.92\% item pairs). Among the dependent item pairs, 83.87\% (26 pairs) showed positive local dependence and 16.13\% showed negative dependence. The positive dependence appeared mostly between the items sharing similar words (e.g., ‘party,’ ‘nervous,’ ‘lively’) and measuring the same content domains. The negative dependence occurred between the items measuring different domains but bespeaking related constructs (e.g., consciousness, apprehension) in reversed wordings. The patterns across the three domains suggested that Neuroticism had the most intra- and inter-domain dependence. Among the 26 item pairs that showed within-domain local dependence, 14 pairs (53.85\%) appeared from the Neuroticism, and all five item pairs that showed inter-domain local item dependence were related to the Neuroticism. The findings suggest that the scale may be in need of additional investigation. The outcomes of the current analysis pointed to possible existence of subscales or confounding traits.

\begin{figure}[h!]
    \begin{center}
    \includegraphics[scale=0.9]{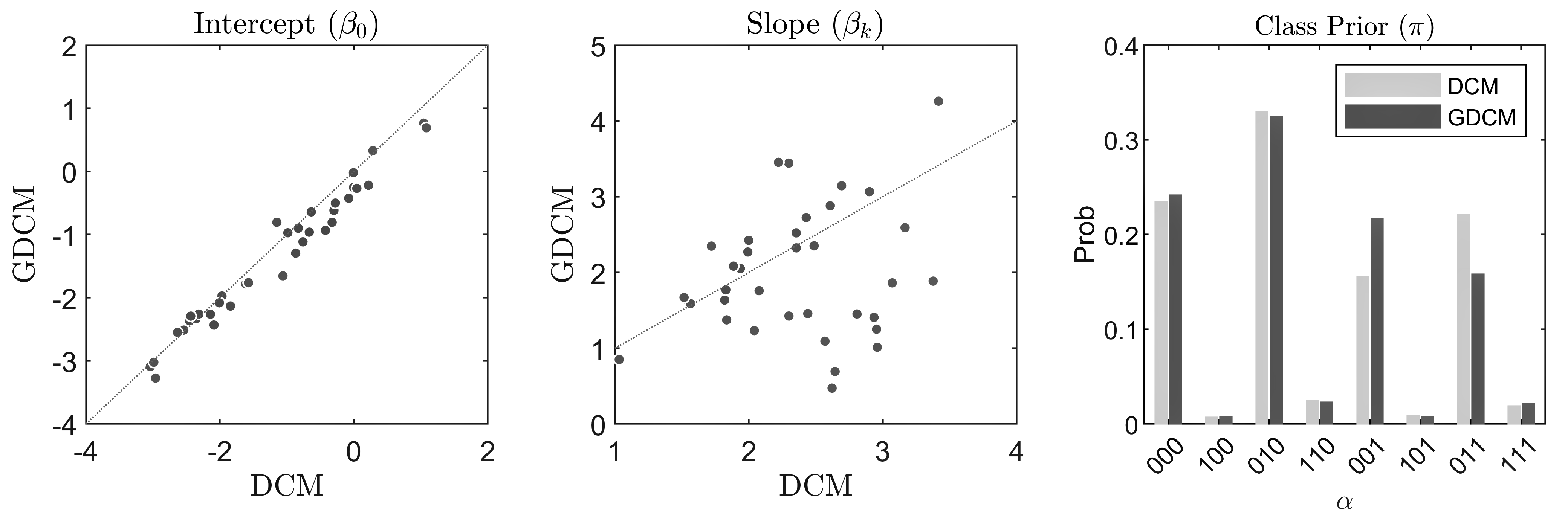}
    \end{center} 
    \vspace{-1em}
    \caption{EPQ-R: Diagnostic Model Parameter Estimates}
    \label{fig:fig6_epq_ipar}

    \vspace{0.5em}
    {\setstretch{0.5} \footnotesize \textit{Note}. $\beta_0$: Item intercept parameter. $\beta_k$: Item slope parameter. $\boldsymbol{\pi}$: Latent class prior.}
\end{figure}

Figure~\ref{fig:fig6_epq_ipar} presents item parameter and class prior probability values that were estimated from the GDCM and DCM. The two models overall showed high consistency in the intercept estimates ($r = .984$) but showed a distinct pattern in the slope estimates ($r = .234$). The most distinction appeared in the Extraversion and Neuroticism domains with average absolute difference, .797 (E) and 1.061 (N) (cf. .298 (P)). While the patterns between the models differed by individual items, it was found that the difference in the slope estimates, combined with the difference in the intercepts generally led to smaller slipping values in DCM. The difference in the item parameter values partially explains the distinct pattern in the class prior estimates. In Figure~\ref{fig:fig6_epq_ipar}, the two models showed distinct patterns in classifying the (001) and (011) groups. Compared to GDCM, DCM yielded a higher prior probability for (011) but a lower probability for (001) ($\hat{\pi}_{011}$ = .221 (DCM) vs. .158 (GDCM); $\hat{\pi}_{001}$ = .156 (DCM) vs. .217 (GDCM)). The attribute classification showed a similar pattern, DCM classifying more individuals into (011) than into (001) (21.00\% vs. 18.69\%; 17.35\% vs. 20.63\% (GDCM)). The distinction in the classification of Extraversion seemed to be related to the difference in the slipping parameter values. Items measuring the Extraversion tended to have smaller slipping values in DCM (on average .158 vs. .288 (GDCM)) and consequently led to relatively higher flagging rate. On the whole, it was observed that when the models disagree in the item parameter values, it led to a distinct pattern in the attribute classification with the DCM resulting in a higher attribute mastery rate.

\section{7. Conclusion}

The study presented an exploratory modeling framework for evaluating local item dependence within the cognitively diagnostic classification models. The framework integrates a discrete Markov network with a generalized DCM so that items’ own interplay can be modeled while performing cognitive diagnosis. Since it explicitly models inter-item dependency, it affords robust inference against probable existence of item interactions. Numerical experimentation with simulated data suggests that the model delivers reliable performance when confronted with locally dependent items. Compared to DCM, GDCM achieved more accurate parameter recovery and showed substantially smaller bias in the item parameter estimates. The GDCM, despite the greater flexibility and generality, did not incur cost of overfitting as it immanently regulated complexity of the item network.

The inferential scheme for GDCM was developed based on pseudo-likelihood and implemented by the EM algorithm. The estimation exploited the coordinate descent algorithm for computational efficiency and optimized for each parameter domain by applying the soft-thresholding and gradient-based Newton iteration. The observations from the Monte Carlo simulation suggest that the estimation performs adequately well, yielding reliable parameter estimates within a few cycles. The study also evaluated fidelity of the standard error estimates and verified that the estimators perform reasonably well.

Throughout, it was substantiated that the suggested framework brings some practical significance. Unlike the regular DCM, GDCM serves dual functioning within a unified framework—cognitive diagnosis and item evaluation. The diagnostic model was formulated in the generalized form so it can support various item-attribute interactions. The network model performs exploratory evaluation so that it can unveil items at risk of local dependence. The experimental application to TIMSS data in particular suggested that GDCM can provide useful precursory information about items’ dependency structure and forestall the cost of overfitting testlet items.

The present work brings a number of future directions. The current simulation study made necessary simplifications to keep the scope of the work at a manageable level. Future studies may verify performance of GDCM in extended settings, such as other item-attribute interactions (e.g., disjunctive, additive, hierarchical attributes), other item response types (e.g., partially-credited, graded response), and more real-life-like data (e.g., missing data, stochastic data, data with covariates). A future study may also relax the condition on $\boldsymbol{Q}$ and investigate functioning of GDCM under modestly misspecified $\boldsymbol{Q}$ entries or extend the model to a fully exploratory scheme so that it can perform cognitive diagnosis while concurrently identifying the item-attribute interaction ($\boldsymbol{Q}$) and item-item interaction ($\boldsymbol{S}$).

\newpage 
\bibliography{GDCM_bib}

\newpage
\section*{Appendix}

\begin{table}[h!]
\renewcommand\thetable{A1}
\centering
\caption{Standard Errors of Item Parameter Estimates: Comparison between GDCM and DCM}
\setlength\tabcolsep{5pt}
\label{tab:atab1_compare_se}
\vspace{-0.5em}

\begin{threeparttable}
{\small
\renewcommand{\arraystretch}{0.8}
\begin{tabular}{L{1.0cm} L{1.0cm} R{0.8cm} R{0.8cm} R{0.8cm} R{0.8cm} R{0.8cm} R{0.8cm} R{0.8cm} R{0.8cm} R{0.8cm} R{0.8cm} R{0.8cm} R{0.8cm}}
\toprule
    &   &\multicolumn{6}{c}{GDCM}   &\multicolumn{6}{c}{DCM} \\ \cmidrule(l){3-8} \cmidrule(l){9-14}
    &   &\multicolumn{2}{c}{LI ($s_{jj^\prime} = 0$)} &\multicolumn{2}{c}{Dyad LD} &\multicolumn{2}{c}{Triad LD} &\multicolumn{2}{c}{LI ($s_{jj^\prime} = 0$)} &\multicolumn{2}{c}{Dyad LD} &\multicolumn{2}{c}{Triad LD} \\ \cmidrule(l){3-4} \cmidrule(l){5-6} \cmidrule(l){7-8} \cmidrule(l){9-10} \cmidrule(l){11-12} \cmidrule(l){13-14}
$s_{jj^\prime}$ 	&$N$	&$\beta_0$	&$\beta_k$	&$\beta_0$	&$\beta_k$	&$\beta_0$	&$\beta_k$	&$\beta_0$ &$\beta_k$	&$\beta_0$ &$\beta_k$	&$\beta_0$ 	&$\beta_k$ \\
\midrule
.5	&500	&.183	&.341	&.172	&.367	&.158	&.393	&.183	&.342	&.172	&.368	&.158	&.395 \\
    &1000	&.128	&.239	&.121	&.258	&.111	&.280	&.128	&.239	&.121	&.258	&.111	&.279 \\
    &3000	&.074	&.136	&.069	&.147	&.064	&.161	&.074	&.136	&.069	&.147	&.064	&.161 \\
1.0	&500	&	    &       &.157	&.397	&.134	&.485	&		&       &.158	&.397	&.134	&.464 \\
    &1000	&       &       &.111	&.281	&.095	&.353	&		&       &.111	&.280	&.094	&.334 \\
    &3000	&       &       &.064	&.162	&.055	&.207	&		&       &.064	&.162	&.054	&.193 \\
\bottomrule
\end{tabular}
}

\vspace{1em}
\begin{tablenotes}
    \linespread{1} \footnotesize
    \item \textit{Note}. $s_{jj^\prime}$: Item network entry for locally dependent items (.5: Moderate dependence, 1.0: Large dependence). $N$: Calibration sample size. LD: Local dependence. All standard errors were evaluated on the Hessian estimates.
\end{tablenotes}
\end{threeparttable}
\end{table}

\begin{table}[h!]
\renewcommand\thetable{A2}
\centering
\caption{$\boldsymbol{\pi}$ and Attribute Profile Recovery: Comparison between GDCM and DCM}
\setlength\tabcolsep{5pt}
\label{tab:atab2_compare_attr}
\vspace{-0.5em}

\begin{threeparttable}
{\small
\renewcommand{\arraystretch}{0.8}
\begin{tabular}{L{0.8cm} L{0.8cm} R{0.8cm} R{0.8cm} R{0.8cm} R{0.8cm} R{0.8cm} R{0.8cm} R{0.8cm} R{0.8cm} R{0.8cm} R{0.8cm} R{0.8cm} R{0.8cm}}
\toprule
    &   &\multicolumn{3}{c}{GDCM}   &\multicolumn{3}{c}{DCM} &\multicolumn{3}{c}{GDCM}   &\multicolumn{3}{c}{DCM} \\ \cmidrule(l){3-5} \cmidrule(l){6-8} \cmidrule(l){9-11} \cmidrule(l){12-14}
$s_{jj^\prime}$	&$N$	&{\footnotesize $\boldsymbol{\pi}$ RMSE} &AAR	&PAR	&{\footnotesize $\boldsymbol{\pi}$ RMSE}	&AAR	&PAR	&{\footnotesize $\boldsymbol{\pi}$ RMSE}	&AAR	&PAR	&{\footnotesize $\boldsymbol{\pi}$ RMSE}	&AAR	&PAR \\
\midrule
    &	&\multicolumn{6}{c}{Null (LI)}	&\multicolumn{6}{c}{} \\ \cmidrule(l){3-8}	 				
0	&500	&.015	&.985	&.944	&.013	&.986	&.946	&		&		&	    &       &       & \\
    &1000	&.011	&.985	&.944	&.011	&.986	&.945	&		&		&	    &       &       & \\
    &3000	&.008	&.986	&.946	&.007	&.986	&.946	&		&		&	    &       &       & \\
    &	&\multicolumn{6}{c}{Dyad Dependence}	&\multicolumn{6}{c}{Triad Dependence} \\ \cmidrule(l){3-8} \cmidrule(l){9-14}
.5	&500	&.014	&.989	&.958	&.013	&.989	&.960	&.014	&.987	&.952	&.013	&.987	&.953 \\
    &1000	&.012	&.989	&.959	&.012	&.989	&.959	&.013	&.988	&.954	&.011	&.988	&.955 \\
    &3000	&.010	&.989	&.960	&.009	&.989	&.960	&.011	&.988	&.955	&.011	&.988	&.956 \\
1.0	&500	&.015	&.986	&.949	&.013	&.986	&.949	&.013	&.979	&.926	&.015	&.978	&.920 \\
    &1000	&.014	&.987	&.953	&.011	&.987	&.951	&.012	&.980	&.927	&.013	&.978	&.921 \\
    &3000	&.012	&.987	&.954	&.012	&.987	&.952	&.012	&.980	&.927	&.012	&.979	&.924 \\
\bottomrule
\end{tabular}
}

\vspace{1em}
\begin{tablenotes}
    \linespread{1} \footnotesize
    \item \textit{Note}. $s_{jj^\prime}$: Item network entry for locally dependent items (.5: Moderate dependence, 1.0: Large dependence). $N$: Calibration sample size. RMSE: Root mean squared error. AAR: Attribute-wise agreement rate. PAR: Pattern-wise agreement rate. LI: Local independence.
\end{tablenotes}
\end{threeparttable}
\end{table}

\newpage
\begin{table}[ht!]
\renewcommand\thetable{A3}
\centering
\caption{TIMSS: Locally Dependent Item Pairs}
\setlength\tabcolsep{5pt}
\label{tab:atab3_timss_ipair}
\vspace{-0.5em}

\begin{threeparttable}
{\small
\renewcommand{\arraystretch}{0.8}
\begin{tabular}{L{1.2cm} L{1.8cm} L{2cm} L{6cm} L{1.2cm}}
\toprule
    &\multicolumn{2}{c}{Item ID}   &    & \\ \cmidrule(l){2-3}  
$\hat{s}_{jj^\prime}$ &Original 	&Calibration 	&Content Domain	& Testlet \\
\midrule
1.037	&M051061A	&20	&Geometric Shapes and Measures	&Y \\
 	&M051061Z	&23	&Geometric Shapes and Measures	&Y \\
.641	&M051061B	&21	&Geometric Shapes and Measures	&Y \\
 	&M051061C	&22	&Geometric Shapes and Measures	&Y \\
.273	&M041046	&10	&Number	                        &   \\
 	&M051236	&24	&Geometric Shapes and Measures	& \\
.155	&M051094	&3	&Number	                        & \\
 	&M051236	&24	&Geometric Shapes and Measures	& \\
.155	&M051140	&1	&Number	                        & \\
 	&M041298	&7	&Number	                        &  \\            
.135	&M051125B	&6	&Data Display	                &Y \\
 	&M041298	&7	&Number	                        & \\ 
.097	&M051060	&4	&Geometric Shapes and Measures	& \\
	&M041298	&7	&Number	                        & \\
.075	&M051094	&3	&Number	                        &  \\      
	&M051060	&4	&Geometric Shapes and Measures	& \\
.071	&M051236	&24	&Geometric Shapes and Measures	& \\
	&M041267	&27	&Geometric Shapes and Measures	& \\
.038	&M051094	&3	&Number	                        & \\
	&M041333	&13	&Geometric Shapes and Measures	& \\
.032	&M051125B	&6	&Data Display	                & Y \\
	&M051125A	&25	&Data Display	                & Y \\
.028	&M051017	&2	&Number	                        &   \\ 
	&M041298	&7	&Number	                        & \\      
.014	&M051129	&5	&Geometric Shapes and Measures	& \\
	&M041298	&7	&Number	                        & \\
.013	&M041177	&15	&Data Display	                & \\
	&M051236	&24	&Geometric Shapes and Measures	& \\
.011	&M041169	&12	&Geometric Shapes and Measures	& \\
	&M051236	&24	&Geometric Shapes and Measures	& \\
.007	&M041007	&8	&Number	                        & \\
	&M051236	&24	&Geometric Shapes and Measures	& \\
.002	&M051017	&2	&Number	                        & \\
	&M051236	&24	&Geometric Shapes and Measures  & \\
\bottomrule
\end{tabular}
}

\vspace{1em}
\begin{tablenotes}
    \linespread{1} \footnotesize
    \item \textit{Note}. Testlet 1: (M051125A, M051125B). Testlet 2: (M051061A, M051061B, M051061C, M051061Z). Testlet 3: (M041276A, M041276B).
\end{tablenotes}
\end{threeparttable}
\end{table}

\newpage
\begin{table}[ht!]
\renewcommand\thetable{A4}
\centering
\caption{TIMSS: Item Parameter Estimates}
\setlength\tabcolsep{5pt}
\label{tab:atab4_timss_ipar}
\vspace{-0.5em}

\begin{threeparttable}
{\footnotesize
\renewcommand{\arraystretch}{0.8}
\begin{tabular}{L{0.5cm} L{1.7cm} R{1cm} R{1cm} R{0.7cm} R{0.7cm} R{1cm} R{1cm} R{0.7cm} R{0.7cm} R{0.7cm} R{0.7cm} R{0.7cm} R{0.7cm}}
\toprule
   & &\multicolumn{8}{c}{Parameter Estimates} &\multicolumn{4}{c}{Standard Error} \\ \cmidrule(l){3-10} \cmidrule(l){11-14}
   & &\multicolumn{4}{c}{GDCM}  &\multicolumn{4}{c}{DCM} &\multicolumn{2}{c}{GDCM}  &\multicolumn{2}{c}{DCM} \\ \cmidrule(l){3-6} \cmidrule(l){7-10} \cmidrule(l){11-12} \cmidrule(l){13-14}
\multicolumn{2}{c}{Item ID}	&$\beta_0$  &$\beta_k$  &$g$    &$s$	&$\beta_0$	&$\beta_k$	&$g$	&$s$	&$\beta_0$	&$\beta_k$	&$g$	&$s$ \\
\midrule
1	&M051140	  &-.364	&1.636	&.410	&.219	&-.302	&1.677	&.425	&.202	&.111	&.174	&.116	&.173 \\
2	&M051017	  &-.203	&1.475	&.450	&.219	&-.211	&1.431	&.447	&.228	&.111	&.170	&.115	&.168 \\
3	&M051094	  &-.677	&2.045	&.337	&.203	&-.708	&2.219	&.330	&.181	&.115	&.181	&.122	&.181 \\
4	&M051060	  &.081	    &1.239	&.520	&.211	&.165	&1.283	&.541	&.190	&.090	&.209	&.093	&.197 \\
5	&M051129	  &.558	    &1.010	&.636	&.172	&.548	&.974	&.634	&.179	&.093	&.216	&.096	&.202 \\
6	&M051125B     &-.365	&1.828	&.410	&.188	&-.274	&1.870	&.432	&.169	&.129	&.182	&.133	&.183 \\
7	&M041298	  &1.342	&.807	&.793	&.104	&1.405	&.974	&.803	&.085	&.145	&.247	&.144	&.234 \\
8	&M041007	  &-.431	&1.355	&.394	&.284	&-.477	&1.350	&.383	&.295	&.113	&.163	&.118	&.163 \\
9	&M041280	  &-.843	&1.240	&.301	&.402	&-.884	&1.229	&.292	&.414	&.120	&.162	&.126	&.164 \\
10	&M041046	 &-.196	    &2.558	&.451	&.086	&-.208	&2.641	&.448	&.081	&.111	&.232	&.115	&.221 \\
11	&M041048	 &-.820	    &1.127	&.306	&.424	&-.838	&1.087	&.302	&.438	&.120	&.161	&.125	&.162 \\
12	&M041169	 &.162	    &1.282	&.540	&.191	&.127	&1.276	&.532	&.197	&.090	&.207	&.092	&.195 \\
13	&M041333	 &-.763	    &1.348	&.318	&.358	&-.787	&1.337	&.313	&.366	&.096	&.181	&.099	&.173 \\
14	&M041262	 &-.749	    &745	&.321	&.501	&-.773	&.734	&.316	&.510	&.129	&.214	&.132	&.212 \\
16	&M041271	 &.172	    &2.163	&.543	&.088	&.121	&2.159	&.530	&.093	&.129	&.212	&.132	&.209 \\
17	&M051111	 &-1.964	&1.614	&.123	&.587	&-2.028	&1.622	&.116	&.600	&.168	&.200	&.179	&.207 \\
18	&M051089	 &-2.553	&2.234	&.072	&.579	&-2.746	&2.373	&.060	&.592	&.213	&.239	&.241	&.263 \\
19	&M051227	 &-3.532	&3.288	&.028	&.561	&-3.891	&3.566	&.020	&.581	&.332	&.349	&.410	&.423 \\
20	&M051061A    &-1.615	&1.816	&.166	&.450	&-1.627	&2.274	&.164	&.344	&.120	&.205	&.124	&.190 \\
21	&M051061B    &-1.080	&3.965	&.253	&.053	&-1.043	&3.577	&.261	&.074	&.100	&.449	&.105	&.282 \\
22	&M051061C    &-1.289	&4.015	&.216	&.061	&-1.264	&4.478	&.220	&.039	&.104	&.419	&.111	&.372 \\
23	&M051061Z    &-5.247	&6.091	&.005	&.301	&-5.520	&5.897	&.004	&.407	&.546	&.580	&.731	&.745 \\
24	&M051236	 &-.525	    &1.001	&.372	&.383	&-.267	&1.136	&.434	&.295	&.091	&.185	&.093	&.176 \\
25	&M051125A    &1.225	    &1.746	&.773	&.049	&1.207	&1.819	&.770	&.046	&.154	&.271	&.157	&.273 \\
26	&M041059	 &.859	    &1.909	&.702	&.059	&.802	&1.914	&.690	&.062	&.121	&.255	&.124	&.246 \\
27	&M041267	 &-.913	    &.835	&.286	&.519	&-.918	&.869	&.285	&.512	&.099	&.177	&.102	&.171 \\
28	&M041276A    &-2.171	&3.388	&.102	&.228	&-2.351	&3.499	&.087	&.241	&.212	&.240	&.234	&.259 \\
29	&M041276B    &-2.235	&2.454	&.097	&.445	&-2.289	&2.462	&.092	&.457	&.217	&.238	&.228	&.247 \\ \cline{1-14}
\multicolumn{2}{c}{Avg}	&-.829	&2.017	&.360	&.276	&-.862	&2.070	&.360	&.275	&.147	&.238	&.162	&.237 \\
\multicolumn{2}{c}{SD}	&1.397	&1.184	&.214	&.176	&1.474	&1.181	&.217	&.180	&.093	&.096	&.128	&.115 \\
\bottomrule
\end{tabular}
}

\vspace{1em}
\begin{tablenotes}
    \linespread{1} \footnotesize
    \item \textit{Note}. $\beta_0$: Item intercept parameter. $\beta_k$: Item slope parameter. $g$: Item guessing parameter. $s$: Item slipping parameter. 
\end{tablenotes}

\end{threeparttable}
\end{table}

\newpage
\begin{table}[ht!]
\renewcommand\thetable{A5}
\centering
\caption{EPQ-R: Locally Dependent Item Pairs}
\setlength\tabcolsep{5pt}
\label{tab:atab5_epq_ipair}
\vspace{-0.5em}

\begin{threeparttable}
{\small
\renewcommand{\arraystretch}{0.7}
\begin{tabular}{L{1.2cm} L{0.8cm} L{0.9cm} L{0.8cm} L{11.5cm}}
\toprule
    &\multicolumn{2}{c}{Item ID}   &    & \\ \cmidrule(l){2-3}  
$\hat{s}_{jj^\prime}$ &Raw 	&{\footnotesize Calib} 	&{\footnotesize Dom}	& Question \\
\midrule
2.556	&42	&18	&E	&Can you easily get some life into a rather dull party? \\
 	&50	&20	&E	&Can you get a party going? \\
1.714	&63	&31	&N	&Would you call yourself a nervous person? \\
 	&74	&35	&N	&Do you suffer from `nerves`? \\
1.608	&35	&14	&E	&Are you rather lively? \\
 	&52	&22	&E	&Do other people think of you as being very lively? \\
1.229	&34	&13	&E	&Are you a talkative person? \\
 	&55	&24	&E	&(R) Are you mostly quiet when you are with other people? \\
.974	&2	&2	&P	&Do you prefer to go your own way rather than act by the rules? \\
 	&30	&12	&P	&(R) Is it better to follow society`s rules than go your own way? \\
.932	&63	&31	&N	&Would you call yourself a nervous person? \\
 	&66	&33	&N	&Would you call yourself tense or `highly-strung`? \\
.787	&63	&31	&N	&Would you call yourself a nervous person? \\
 	&64	&32	&N	&Are you a worrier? \\
.784	&64	&32	&N	&Are you a worrier? \\
 	&73	&34	&N	&Do you worry too long after an embarrassing experience? \\
.684	&37	&16	&E	&Do you enjoy meeting new people? \\
 	&44	&19	&E	&Do you like mixing with people? \\
.669	&66	&33	&N	&Would you call yourself tense or `highly-strung`? \\
 	&74	&35	&N	&Do you suffer from `nerves`? \\
.508	&42	&18	&E	&Can you easily get some life into a rather dull party? \\
 	&52	&22	&E	&Do other people think of you as being very lively? \\
.466	&64	&32	&N	&Are you a worrier? \\
 	&74	&35	&N	&Do you suffer from `nerves`? \\
.314	&44	&19	&E	&Do you like mixing with people? \\
 	&51	&21	&E	&Do you like plenty of bustle and excitement around you? \\
.312	&2	&2	&P	&Do you prefer to go your own way rather than act by the rules? \\
 	&15	&6	&P	&(R) Do you take much notice of what people think? \\
.304	&56	&25	&N	&Does your mood often go up and down? \\
 	&57	&26	&N	&Do you ever feel `just miserable` for no reason? \\
.261	&62	&30	&N	&Are you often troubled about feelings of guilt? \\
 	&73	&34	&N	&Do you worry too long after an embarrassing experience? \\
.227	&62	&30	&N	&Are you often troubled about feelings of guilt? \\
 	&64	&32	&N	&Are you a worrier? \\
.223	&59	&27	&N	&Are you an irritable person? \\
 	&66	&33	&N	&Would you call yourself tense or `highly-strung`? \\
.220	&60	&28	&N	&Are your feelings easily hurt? \\
 	&64	&32	&N	&Are you a worrier? \\
.209	&34	&13	&E	&Are you a talkative person? \\
 	&52	&22	&E	&Do other people think of you as being very lively? \\
.166	&34	&13	&E	&Are you a talkative person? \\
 	&35	&14	&E	&Are you rather lively? \\
.118	&36	&15	&E	&Can you usually let yourself go and enjoy yourself at a lively party? \\
 	&51	&21	&E	&Do you like plenty of bustle and excitement around you? \\
.055	&60	&28	&N	&Are your feelings easily hurt? \\
 	&73	&34	&N	&Do you worry too long after an embarrassing experience? \\
.054	&35	&14	&E	&Are you rather lively? \\
 	&41	&17	&E	&Do you usually take the initiative in making new friends? \\
.050	&64	&32	&N	&Are you a worrier? \\
 	&66	&33	&N	&Would you call yourself tense or `highly-strung`? \\
.037	&63	&31	&N	&Would you call yourself a nervous person? \\
 	&75	&36	&N	&Do you often feel lonely? \\
\bottomrule
\end{tabular}
}

\vspace{1em}
\begin{tablenotes}
    \linespread{1} \footnotesize
    \item \textit{Note}. Raw: ID from the original data. Calib: ID from the calibration data. Dom: Domain.
\end{tablenotes}
\end{threeparttable}
\end{table}

\newpage
\begin{table}[ht!]
\renewcommand\thetable{A5}
\centering
\caption{EPQ-R: Locally Dependent Item Pairs, Continued}
\setlength\tabcolsep{5pt}
\vspace{-0.5em}

\begin{threeparttable}
{\small
\renewcommand{\arraystretch}{0.7}
\begin{tabular}{L{1.2cm} L{0.8cm} L{0.9cm} L{0.8cm} L{11.5cm}}
\toprule
    &\multicolumn{2}{c}{Item ID}   &    & \\ \cmidrule(l){2-3}  
$\hat{s}_{jj^\prime}$ &Raw 	&{\footnotesize Calib} 	&{\footnotesize Dom}	& Question \\
\midrule
-.055	&30	&12	&P	&(R) Is it better to follow society`s rules than go your own way? \\
 	&73	&34	&N	&Do you worry too long after an embarrassing experience? \\
-.129	&42	&18	&E	&Can you easily get some life into a rather dull party? \\
 	&73	&34	&N	&Do you worry too long after an embarrassing experience? \\
-.148	&15	&6	&P	&(R) Do you take much notice of what people think? \\
 	&62	&30	&N	&Are you often troubled about feelings of guilt? \\
-.224	&15	&6	&P	&(R) Do you take much notice of what people think? \\
 	&73	&34	&N	&Do you worry too long after an embarrassing experience? \\
-.430	&15	&6	&P	&(R) Do you take much notice of what people think? \\
 	&60	&28	&N	&Are your feelings easily hurt? \\
\bottomrule
\end{tabular}
}

\vspace{1em}
\begin{tablenotes}
    \linespread{1} \footnotesize
    \item \textit{Note}. Raw: ID from the original data. Calib: ID from the calibration data. Dom: Domain.
\end{tablenotes}
\end{threeparttable}
\end{table}

\newpage
\begin{table}[ht!]
\renewcommand\thetable{A6}
\centering
\caption{EPQ-R: Item Parameter Estimates}
\setlength\tabcolsep{5pt}
\label{tab:atab6_epq_ipar}
\vspace{-0.5em}

\begin{threeparttable}
{\footnotesize
\renewcommand{\arraystretch}{0.8}
\begin{tabular}{L{0.8cm} R{1cm} R{1cm} R{0.7cm} R{0.7cm} R{1cm} R{1cm} R{0.7cm} R{0.7cm} R{0.7cm} R{0.7cm} R{0.7cm} R{0.7cm}}
\toprule
   &\multicolumn{8}{c}{Parameter Estimates} &\multicolumn{4}{c}{Standard Error} \\ \cmidrule(l){2-9} \cmidrule(l){10-13}
   &\multicolumn{4}{c}{GDCM}  &\multicolumn{4}{c}{DCM} &\multicolumn{2}{c}{GDCM}  &\multicolumn{2}{c}{DCM} \\ \cmidrule(l){2-5} \cmidrule(l){6-9} \cmidrule(l){10-11} \cmidrule(l){12-13}
ID	&$\beta_0$  &$\beta_k$  &$g$    &$s$	&$\beta_0$	&$\beta_k$	&$g$	&$s$	&$\beta_0$	&$\beta_k$	&$g$	&$s$ \\
\midrule
1	&-2.513	&3.067	&.075	&.365	&-2.539	&2.901	&.073	&.411	&.139	&.270	&.142	&.255 \\
2	&-.218	&1.863	&.446	&.162	&.219	&3.070	&.555	&.036	&.076	&.431	&.074	&.566 \\
3	&-2.364	&2.352	&.086	&.503	&-2.454	&2.487	&.079	&.492	&.131	&.258	&.137	&.250 \\
4	&-.971	&1.613	&.275	&.345	&-.985	&1.548	&.272	&.363	&.082	&.248	&.083	&.233 \\
5	&-3.091	&2.270	&.043	&.694	&-3.039	&1.993	&.046	&.740	&.180	&.301	&.177	&.297 \\
6	&-.805	&.851	&.309	&.489	&-1.150	&1.033	&.241	&.529	&.086	&.241	&.086	&.226 \\
7	&-2.329	&1.589	&.089	&.677	&-2.352	&1.565	&.087	&.687	&.129	&.271	&.131	&.261 \\
8	&-2.259	&2.324	&.095	&.484	&-2.315	&2.355	&.090	&.490	&.125	&.255	&.129	&.246 \\
9	&-2.549	&1.635	&.073	&.714	&-2.632	&1.820	&.067	&.693	&.141	&.284	&.148	&.270 \\
10	&-1.973	&1.669	&.122	&.575	&-1.969	&1.516	&.123	&.611	&.112	&.251	&.113	&.242 \\
11	&-3.022	&2.880	&.046	&.536	&-2.986	&2.607	&.048	&.594	&.174	&.283	&.173	&.274 \\
12	&-.806	&1.425	&.309	&.350	&-.325	&2.299	&.419	&.122	&.076	&.297	&.075	&.328 \\
13	&-.961	&1.452	&.277	&.380	&-.666	&2.807	&.339	&.105	&.112	&.180	&.115	&.187 \\
14	&-.617	&1.886	&.350	&.219	&-.298	&3.375	&.426	&.044	&.113	&.219	&.110	&.247 \\
15	&-.018	&2.523	&.495	&.076	&-.007	&2.353	&.498	&.087	&.105	&.212	&.109	&.194 \\
16	&.765	&3.144	&.682	&.020	&1.042	&2.693	&.739	&.023	&.128	&.483	&.124	&.325 \\
17	&-.640	&2.052	&.345	&.196	&-.637	&1.936	&.346	&.214	&.110	&.162	&.114	&.159 \\
18	&-3.274	&1.251	&.036	&.883	&-2.960	&2.954	&.049	&.502	&.200	&.229	&.252	&.267 \\
19	&.692	&2.725	&.666	&.032	&1.082	&2.428	&.747	&.029	&.131	&.438	&.125	&.298 \\
20	&-2.081	&1.458	&.111	&.651	&-2.008	&2.440	&.118	&.394	&.157	&.194	&.169	&.192 \\
21	&-.253	&1.770	&.437	&.180	&-.005	&1.829	&.499	&.139	&.105	&.175	&.109	&.170 \\
22	&-1.653	&1.407	&.161	&.561	&-1.059	&2.934	&.258	&.133	&.119	&.175	&.125	&.183 \\
23	&-1.781	&4.261	&.144	&.077	&-1.614	&3.416	&.166	&.142	&.149	&.230	&.146	&.196 \\
24	&-1.115	&2.591	&.247	&.186	&-.760	&3.166	&.319	&.083	&.114	&.209	&.117	&.202 \\
25	&-.268	&3.444	&.434	&.040	&.044	&2.298	&.511	&.088	&.094	&.316	&.093	&.208 \\
26	&-.425	&2.348	&.395	&.127	&-.077	&1.721	&.481	&.162	&.095	&.199	&.093	&.170 \\
27	&-1.762	&2.083	&.147	&.420	&-1.573	&1.885	&.172	&.423	&.131	&.169	&.123	&.163 \\
28	&.330	&1.760	&.582	&.110	&.285	&2.077	&.571	&.086	&.095	&.199	&.094	&.209 \\
29	&-1.293	&3.454	&.215	&.103	&-.868	&2.222	&.296	&.205	&.114	&.206	&.102	&.165 \\
30	&-.899	&1.374	&.289	&.383	&-.830	&1.835	&.304	&.268	&.100	&.152	&.101	&.156 \\
31	&-2.433	&.473	&.081	&.876	&-2.087	&2.621	&.110	&.370	&.129	&.179	&.149	&.184 \\
32	&-.932	&1.013	&.282	&.480	&-.424	&2.959	&.396	&.073	&.099	&.174	&.095	&.223 \\
33	&-2.291	&1.093	&.092	&.768	&-2.435	&2.568	&.081	&.467	&.137	&.177	&.171	&.201 \\
34	&-.501	&1.233	&.377	&.325	&-.276	&2.041	&.431	&.146	&.097	&.166	&.094	&.176 \\
35	&-2.262	&.693	&.094	&.828	&-2.141	&2.644	&.105	&.377	&.128	&.175	&.152	&.186 \\
36	&-2.132	&2.424	&.106	&.427	&-1.842	&1.999	&.137	&.461	&.151	&.185	&.136	&.172 \\ \cline{1-13}
Avg	&-1.353	&1.985	&.250	&.396	&-1.185	&2.344	&.283	&.300	&.121	&.239	&.125	&.230 \\
SD	&1.080	&.845	&.180	&.257	&1.160	&.564	&.202	&.219	&.028	&.079	&.035	&.075 \\
\bottomrule
\end{tabular}
}

\vspace{1em}
\begin{tablenotes}
    \linespread{1} \footnotesize
    \item \textit{Note}. $\beta_0$: Item intercept parameter. $\beta_k$: Item slope parameter. $g$: Item guessing parameter. $s$: Item slipping parameter.
\end{tablenotes}

\end{threeparttable}
\end{table}

\end{document}